\documentstyle[emulateapj_h,11pt,eqsecnum]{article}
\input epsf
\newcommand{\bq}{\begin{equation}}
\newcommand{\eq}{\end{equation}}
\newcommand{\bqr}{\begin{eqnarray}}
\newcommand{\eqr}{\end{eqnarray}}
\newcommand{\nm}{\nonumber}

\def\balpha{\mbox{\boldmath$\alpha$}}
\def\bbeta{\mbox{\boldmath$\beta$}}
\def\btheta{\mbox{\boldmath$\theta$}}
\def\pderiv(#1/#2){\mathchoice{\partial#1\over\partial#2}
    {\partial#1/\partial#2} {\partial#1/\partial#2}
    {\partial#1/\partial#2}}
\def\<#1>{\langle{#1}\rangle}
\def\quo#1/#2{\mathchoice  {\hbox{$#1\over#2$}} {{#1\over#2}}
                {\scriptstyle{#1\over#2}}
                {#1\mskip-1.5mu/\mskip-1.5mu#2}}
\def\invh50{h_{50}^{-1}}
\def\msolar{M_\odot}
\def\half{{\quo1/2}}
\def\arcsec{^{\prime\prime}}

\def\etal{et~al.\ }

\def\figureps[#1,#2]#3.{\bgroup\vbox{\epsfxsize=#2
    \hbox to \hsize{\hfil\epsfbox{#1}\hfil}}\vskip12pt
    \small\noindent Figure#3. \def\par{\endgraf\egroup\vskip12pt}}

\def\hfigureps[#1,#2,#3]#4.{
    \bgroup\vbox{\centerline{\epsfxsize=#3\epsfbox{#1}\hfil
              \epsfxsize=#3\epsfbox{#2}}} \small\noindent Figure#4.
    \def\par{\endgraf\egroup\vskip12pt}}

\begin{document}
\renewcommand{\thefootnote}{\fnsymbol{footnote}}
\title{\bf Non-parametric Reconstruction of Abell 2218\footnotemark\ 
from Combined Weak and Strong Lensing} \footnotetext{Based on
observations made with the NASA/ESA Hubble Space Telescope, obtained
from the data archive at the Space Telescope Science Institute. STScI
is operated by the association of universities for research in
Astronomy, Inc. under the NASA contract N-AS 5-26555.}
{\vskip 10pt}
\author{Hanadi M. AbdelSalam, Prasenjit Saha} \affil{Department of
Physics (Astrophysics), Keble Rd., OX1 3RH, Oxford, UK.}
\and 
\author{Liliya L.R. Williams} \affil{Institute of
Astronomy, Madingley Rd., CB3 0HA, Cambridge, UK.}
\date{\today} 

\begin{abstract}
We apply a new non-parametric technique to reconstruct, with
uncertainties, the projected mass distribution of the inner region of
Abell 2218, using combined strong and weak lensing constraints from
multiple-image systems and arclets with known redshifts.  The
reconstructed mass map broadly resembles previous, less detailed,
parametric models, but when examined in detail shows several
sub-structures, not necessarily associated with light but strongly
required by the lensing data.  In particular, the highest mass peak is
offset by $\sim30h^{-1}_{50}\,\rm kpc$ from the main light peak, and
projected mass-to-light in the directions of different cluster
galaxies varies by at least a factor of 10.  On comparing with mass
estimates from models of the X-ray emitting gas, we find that the
X-ray models under-predict the enclosed mass profile by, {\it at
least\/} a factor of 2.5; the discrepancy gets worse if we assume that
mass traces light to the extent allowed by the lensing constraints.
\end{abstract}


\section{Introduction}

Galaxy clusters are thought to be the largest and most recently
assembled gravitationally bound entities in the universe and, thus, an
accurate determination of their masses is of utmost importance to full
understanding of the formation and evolution of cosmic structures,
mapping of dark matter in large scales and for constraining
cosmological parameters. Traditional methods for inferring mass
distributions in clusters are (i) dynamical methods, in which the
observed line-of-sight velocity distribution of the luminous cluster
galaxies is used in conjunction with the virial theorem, and (ii)
X-ray methods, in which the investigation of the X-ray emission of
the intracluster hot gas is used to trace the cluster potential. Both
these traditional methods depend on restrictive assumptions about the
geometrical and dynamical state of clusters.  A more sophisticated
method, which proved to be very reliable and independent of any prior
assumptions, is gravitational lensing.

Theoretically, the above three methods have to yield the same cluster
masses if the clusters are dynamically relaxed, but recent literature
shows a lot of discussion of and disagreement of cluster masses
derived from observed velocity dispersions, observations of X-ray
emitting gas and gravitational lensing (Miralda-Escud\'e and Babul
1995, Wu and Fang 1997). Early studies based upon few selected
clusters claimed a mass estimate discrepancy of at least $\sim 2-3$
among them. This discrepancy can well be attributed to projection
effects, non-thermal pressure or the specific cluster is still in the
formation era. Evidently, the problem is relevant to the precise
determination of mass distribution and to the dynamical evolution of
those clusters. For clusters which are still in an ongoing merging
phase, i.e., not in hydrostatic equilibrium, the hot gas does not
follow the gravitational potential of the cluster and thereby the
X-ray cluster mass is uncertain and should be different from the
gravitational lensing-derived mass and/or even the virial mass. 

Within gravitational lensing, there are two distinct methods for
inferring masses from lensing observations; (i) the parametric
model-fitting method for the strong lensing region (giant arcs and
multiple images), and (ii) statistical distortion method for the weak
lensing regions (weakly distorted single arclets). The parametric
model-fitting starts from the inner most regions of galaxy cluster,
where lensing information is richest and extends outwards. It requiers
fitting parametrized profiles to cluster-sized galaxies (Kneib \etal
96 [hereafter KESCS96]). It works efficiently with one or two multiple
image systems but becomes difficult to implement with several multiple
image systems spanning outwards and hence the fitted model becomes
non-unique. On the other hand, the statistical distortions method
starts from regions where the distortions are very tiny and weak, i.e
outer regions of galaxy clusters, and goes inwards. It requires
averaging statistically the ellipticities of background galaxies over
patches of sky, typically $\sim 20^{\prime\prime}$
in size, to trace the cluster mass
distribution from its shear field (Kaiser and Squires 1993 [hereafter
KS93], Kaiser 1995, Schneider and Seitz 1995). Methods, which are
based on KS93 algorithm, suffer from a global invariance
transformation known as the ``mass-sheet-degenracy'' and also cannot
be used to probe the inner regions of rich clusters, because the
lensing properties change very rapidly over that sampling scale.

In intermediate regions, with no occurrence of
multiple images but having highly distorted arclets, neither of the above
methods can be used. However, several attempts to reonstruct the mass
distribution in this region were made (Kaiser 95, Schneider and Seitz
1995, Seitz \& Schneider 1996), but their formulation of the problem
made it non-linear and still suffer from the so called mass-sheet
degeneracy and boundary effects.

Obviously, the problem of mapping the cluster mass distribution from
gravitational lensing require an independent non-parametric method
that can be used simultaneously in regions with varying lensing
strength. The required method should disentangle the systematic
effects that plagued the existing methods encoded in forms of
non-uniqueness.

In this paper we present a new and general reconstruction technique
that combines regions with varying lensing strength the strong and
overcomes the drawbacks of the existing methods. The technique
is basically an extension of the non-parametric cluster inversion
described by AbdelSalam \etal (1998) [hereafter Paper I] for the strong
lensing regime to incorporate the extra information encoded in the
observations of single/weakly distorted arclets. The extension
results in a method that combines strong and
weak lensing data in a mass reconstruction and overcomes the mass
sheet degeneracy. In general, the method is non-parametric and similar
to the weak lensing (or statistical distortions) method of KS93, but
our formulation of the problem is manifestly linear in all the regimes
and hence the problem becomes simpler and can thereby easily
include the strong lensing regime.

We apply the technique to the spectacular cluster-lens A2218 to
reconstruct its projected mass distribution, with uncertainties, from
combined strong and weak lensing.  This cluster has previously been
studied through parametric lens modeling of strong lensing and arcs by
Kneib \etal (1995) and KESCS96, non-parametric inversion from
statistical distortions by Squires \etal (1996), and parametric
modeling of statistical distortions by Smail \etal (1997).

\section{The Reconstruction Method}

The method we will follow here is a free-form or non-parametric one
that reconstructs a pixellated mass distribution.  It is basically the
same as in Paper~I, except that (a)~in this work we implement
constraints from weak as well as strong lensing whereas the earlier
work implemented only strong lensing and described the extension to
weak lensing, and (b)~this paper uses Gaussian pixels whereas Paper~I
used square pixels---the difference is tiny for the results of either
paper.  Since Paper~I already has a full description, we will only
summarize the technique here.

\subsection{Pixellization of the mass distribution}

The lens plane is divided into $N\times N$ pixels with inter-pixel
distance $a$.  The $mn$th pixel is a Gaussian tent with `dispersion'
$a/2$ and peak height $\kappa_{mn}$. That is to say, if the $mn$-th
pixel is centered at $\btheta_{mn}$ it has a mass profile
\begin{equation}
\kappa_{mn}\exp\left(-2(\btheta-\btheta_{mn})^2\over a^2\right).
\label{pixprof}
\end{equation}
Hereafter, we refer to $a$ as the pixel size.  We measure
$\kappa_{mn}$ in units of the critical density $\Sigma_{\rm crit}$ for
sources at infinity.  Thus the total mass is
\begin{equation}
M_{\rm total} = \frac{a^2\pi}{2}\Sigma_{\rm crit}\sum_{mn}\kappa_{mn}.
\label{mass}
\end{equation}
The appropriately scaled arrival time of a light ray from a background
source at an unlensed angular position $\bbeta$ via a point $\btheta$
in the pixellated lens plane to the observer is
\begin{equation}
\tau(\btheta) = \frac{1}{2}(\btheta-\bbeta)^2-
\frac{D_{ds}}{D_s}\sum_{mn}\kappa_{mn}\psi_{mn}(\btheta),
\label{tau}
\end{equation}
where
\begin{equation}
\psi_{mn}(\btheta)=\frac{1}{\pi}\int
\exp\left(-2(\btheta'-\btheta_{mn})^2\over a^2\right)
\ln|\btheta-\btheta^\prime|\,d^2\btheta'.
\label{pixint}
\end{equation}
Thus, $\kappa_{mn}\psi_{mn}(\btheta)$ represents the contribution of
the $mn$-th pixel to the total gravitational potential of the lens.
Its derivatives with respect to $\btheta$ represent contributions to
the bending angle and amplification components.

See Appendix A for explicit expressions for the function $\psi_{mn}$
and its derivatives.

\subsection{Lensing observables and constraint equations}
The basic approach of our technique is that the lens equation
($\nabla\tau(\btheta)=0$) and shape parameters or distortions of
gravitationally lensed images ($\nabla\nabla\tau(\btheta)$), all at
$\btheta$ values corresponding to the observed image locations, are
considered as rigid linear constraints on the mass distribution.

Writing the lens equation at an observed image location $\btheta_1$
\begin{equation}
\bbeta=\btheta_1-\frac{D_{ds}}{D_s}
\sum_{mn}\kappa_{mn}\nabla_{\btheta}\psi_{mn}(\btheta_1),
\label{lens}
\end{equation}
each image supplies us with two-component constraint equation.  But we
have to solve for the unknown source position $\bbeta$. Thus the
number of constraints on the mass distribution from multiple images is
$\rm 2(\langle images\rangle-\langle sources \rangle)$.  These strong
lensing constraints are linear equality constraints.

{\vskip 10pt}
\figureps[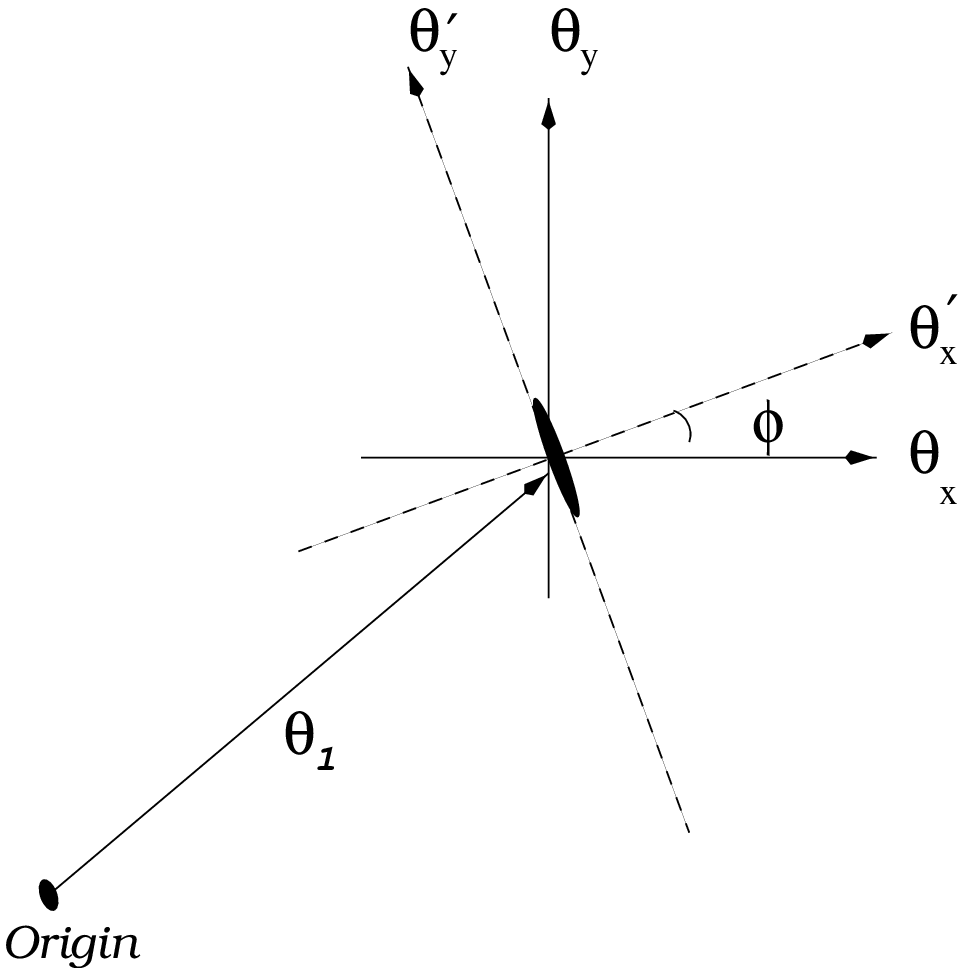,.8\hsize] 1. Illustration of the notation used
in the text for explaining the constraints from weak lensing. The
$\theta_{x'},\theta_{y'}$ axes are the principal axes of the observed
stretched arclet, and are rotated by an observed angle $\phi$ with
respect to the main coordinate axes $\theta_x,\theta_y$.


The constraints from the observed shape parameters of distorted images
(arclets) can be obtained from the the Hessian of the arrival time
function.  If an arclet at $\btheta_1$ is observed to be stretched
along the direction $\btheta_{y'}$ by at least a factor of $\epsilon$
compared to the perpendicular direction $\btheta_{x'}$ (see
Fig.~1), then by considering the
$\theta_{x'},\theta_{y'}$ components of the inverse amplification
matrix we have
\begin{equation}
\epsilon\left|\frac{\partial^2}{\partial{\theta_{x'}}^2}
\tau(\btheta_1)\right|\le 
\left|\frac{\partial^2}{\partial{\theta_{y'}}^2}
\tau(\btheta_1)\right|.
\label{ineq}
\end{equation}
If we can independently infer the image parity we can remove the
absolute value signs in equation (\ref{ineq}), which will then supply
us with a linear constraint equation on the $\kappa_{mn}$, because
rotation to $\theta_{x'},\theta_{y'}$ coordinates is a linear
transformation.

While constraints of the form (\ref{ineq}) can be used for arclets in
multiply-imaged systems (and we do so), they are most useful for
singly imaged arclets (weak lensing).  For statistical distortions the
form of the constraint stays exactly the same. If the shear field is
known accurately enough the inequality can become an equality, and can
be supplemented by the additional constraint
\begin{equation}
\frac{\partial^2}{\partial\theta_{x'}\partial\theta_{y'}}
\tau(\btheta_1) = 0.
\end{equation}

In our work the weak lensing constraints are expressed in implicit
form (Eq.\ \ref{ineq}), whereas in previous work (e.g.\ KS93, Kaiser
1995, Seitz \& Schneider 1996) it is usual to express these in
explicit form.  The advantage of the implicit form is that the
constraints remain linear in all lensing regimes.

\subsection{Breaking the mass sheet degeneracy}
The mass sheet degeneracy is as follows.  Consider the lens equation in
the form
\begin{equation}
\bbeta = \btheta - \frac{D_{ds}}{D_s}\balpha.
\label{lenseq}
\end{equation}
The transformation
\begin{equation}
\balpha \rightarrow r\balpha +
\frac{D_s}{D_{ds}}(1-r)\btheta,   \qquad
\bbeta \rightarrow r\bbeta
\label{sheet}
\end{equation}
simply amounts to multiplying both sides of Eq.~(\ref{lenseq}) by $r$.
Clearly, image positions will be unaffected by such a
transformation. Also, since the magnification matrix is just
${\partial\btheta/\partial\bbeta}$ the transformation will just
rescale the whole matrix by $r^{-1}$; in other words, relative shear and
magnifications will be unaffected but absolute amplification will
change.  Physically, the transformation (\ref{sheet}) amounts to
rescaling the mass everywhere and then adding a constant mass sheet;
hence the name mass-sheet degeneracy.

It is sometimes thought that measuring absolute amplifications is the
only way to break the mass-sheet degeneracy. This is not so.  It is
easy to verify that the transformation (\ref{sheet}) will not have the
same invariance property for two different source redshifts at the
same time.  Thus lensing data with more than one source redshift
breaks the mass-sheet degeneracy.

{\vskip 10pt}
\subsection{Mass Profile Reconstruction}

The above lensing constraints, together with the physical
requirement of $\kappa_{mn}\ge 0$, by themselves are insufficient to
constrain the lens uniquely. The fact that the number of lensing
constraints is far less than the number of pixels leaves a vast family
of mass distributions, all perfectly consistent with the lensing
observations.  It is now necessary to add more information, based on
some criteria for physical plausibility.

As discussed in Paper~I, it is particularly interesting  to consider
mass distributions that minimize
\begin{equation}
\sum_{mn}\left[\left(\sum_{m'n'}\kappa_{m'n'}\right)L_{mn} -
\kappa_{mn}\right]^2 + \sigma^4\sum_{mn}\left(\nabla^2\kappa_{mn}\right)^2
\label{reg}
\end{equation}
where $L_{mn}$ represents the light profile normalized to unit total
luminosity.  In Eq.\ (\ref{reg}), the first term tends to minimize the
$M/L$ variations while the second term smoothes the mass map on scales
of $\le \sigma$.  Such minimization may be considered as
regularization with respect to the light distribution and the
smoothing scale $\sigma$, and is readily implemented via standard
numerical algorithms, such as the NAG routine E04NFF.  It serves two
purposes: (i) as a test whether light is indeed a fair or biased
tracer of mass, and (ii) a basis for Monte-Carlo simulations to
estimate uncertainties in the mass map.  We go into details in the
next section, but for now we emphasize strongly that our technique
considers the light distribution as a secondary information
subservient to the rigid constraints from the observed lensing data.

\section{Lensed Observables in Abell 2218}

Abell 2218 is an exceptionally rich lensing cluster, at redshift
$z=0.175$, which hosts 7 multiple image systems and over 100 arclets.
In the present paper, we will use all the secure informations that
lensing provides in this first application of combined strong and weak
lensing. Our analysis of A2218 is based on the archival WFPC2/HST
images and we take the redshifts of the resolved images from Ebbels
\etal (1998).

Throughout this paper we use $\Omega=1;\ \Lambda=0;\ H_0= 50h$
km/s/Mpc which corresponds to an angular size of $1''=3.84\,\rm kpc$
at the redshift of Abell 2218.

To refer to individual cluster and background galaxies, we will follow
the three-digit numbering scheme of Le Borgne \etal (1992); but for
objects referred to often we adopt simpler names below.

\subsection{Multiply Lensed features}

Of the multiply imaged background galaxies, we use the three that have
secure spectroscopic redshifts and one with a photometric redshift.
Positions and redshifts are listed in Table 1 and plotted in Fig.~2.
\placetable{tbl-1}
{\vskip 10pt}
{\small
\begin{center}
\begin{tabular}{cccrr}\tableline
  ID   & $z$   & Image & $x$\hfil\null & $y$\hfil\null \\ 
\tableline
  384  & 2.515 &  a1   & $ 14.5$ & $ 19.1$ \\     
       & 2.515 &  a2   & $ 17.0$ & $ 15.4$ \\ 
\tableline
  328  & 0.702 &  b1   & $-14.7$ & $ 16.1$ \\  
  389  & 0.702 &  b2   & $-17.8$ & $-15.4$ \\ 
\tableline
  289  & 1.034 &  c1   & $-61.7$ & $  7.2$ \\
       & 1.034 &  c2   & $-61.5$ & $  0.6$ \\ 
\tableline
  730  & 1.1   &  d1   & $-75.2$ & $ -1.4$ \\
       & 1.1   &  d2   & $-74.6$ & $ -6.0$ \\
       & 1.1   &  d3   & $-73.6$ & $ -9.0$ \\ 
\tableline
\end{tabular}
\end{center}
Table 1: Positions of various sets of multiple images ($x,y$ in
arcsec), as taken from the HST image, used in our
reconstruction.
{\vskip 10pt}
}

\figureps[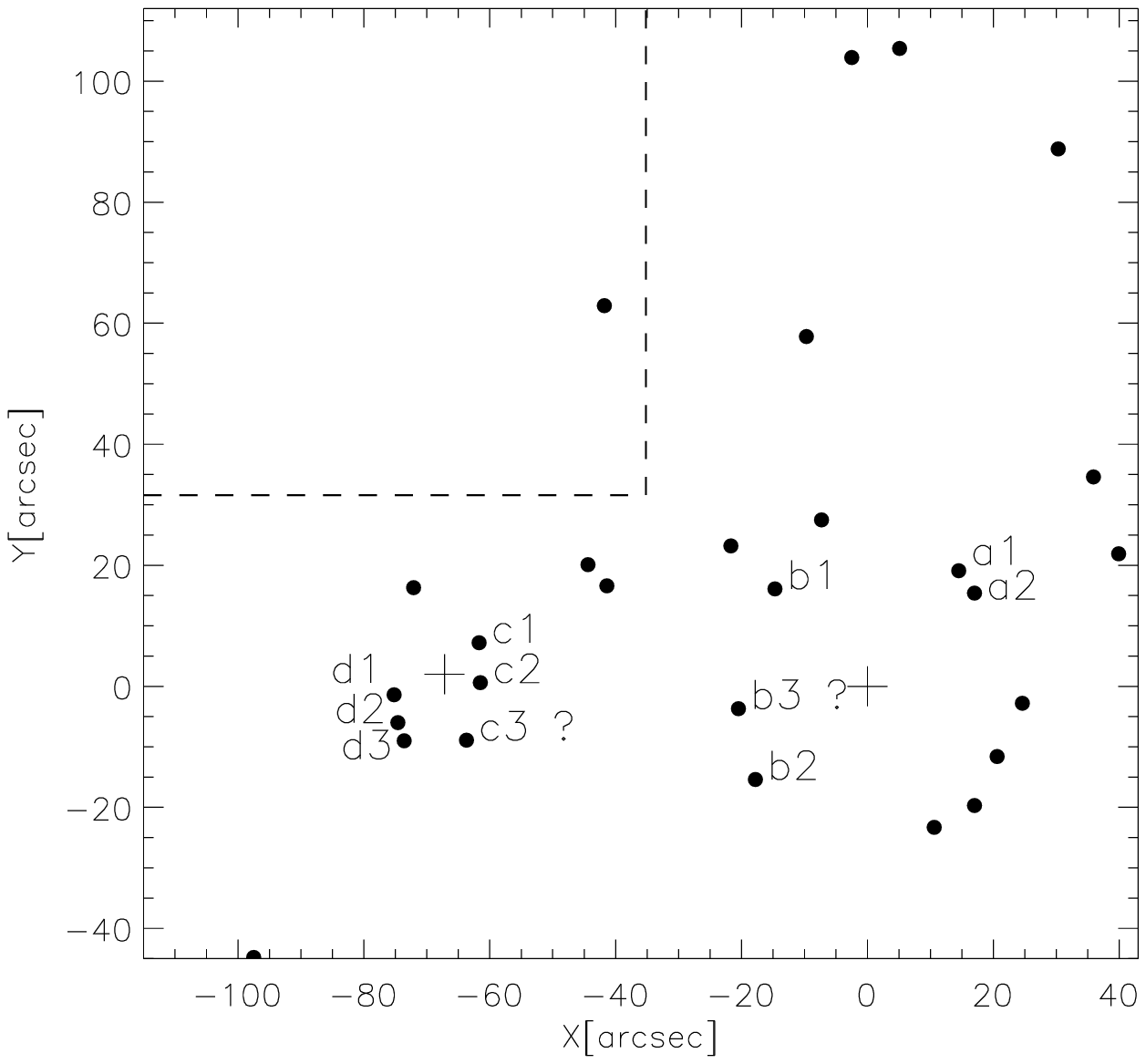,\hsize] 2. The positions of the multiple
images and the arclets in the WFPC2/HST.

{\vskip 10pt}
$\bullet$ {\it The `a' system ($z=2.515$):}\\ The spectacular arc
system \#384 clearly shows an internally symmetric pattern of
unresolved knots which characterizes the system as a fold arc
comprising two merging images (which we will call a1 \& a2) with
opposite parities.  It is identified to be an image of a background
star forming galaxy, with the visible knots representing the H II
regions.
{\vskip 10pt}
$\bullet$ {\it The `b' system ($z=0.702$):}\\ The identification of
the multiple images in this system is somewhat complicated. The red
arc \#359 was initially identified by Pello \etal (1992) to be an
image of a background spheroidal galaxy. It was later interpreted by
Kneib \etal (1995) using a ground-based image to be two merging images
forming a fold arc with \#328 as a counter image. However, the
improved resolution of HST revealed that such a configuration for the
arc \#359 cannot be true (KESCS96) since a
faint extension connecting it to \#337 is now revealed, thus strongly
suggesting it as a counter image too. On the other hand, \#337 and
\#389 yield a similar color to \#359, thus indicating that \#389 is
another counter image of \#359. Since such a system is complicated and
no simple model can explain it, we tried various input
configurations. The only plausible solutions can be obtained when
\#328 and \#389 (which we call b1 and b2) are considered as two images
from the same source, and \#359 (which we call `b3?') is an arclet
arising from another source, possibly a different component of the
same lensed galaxy.
{\vskip 10pt}
$\bullet$ {\it The `c' system ($z=1.034$):}\\ The blue arc \#289 has a
large amount of internal structure in the southern part, which is
highly magnified but appears to be slightly distorted and thus
characterized to be singly imaged while the tail visible on the north
is is highly elongated with tiny breaks and definitely multiply
imaged. Nevertheless, this complex configuration can be explained by a
background source lying near a cusp caustic but with a high portion of
it being outside the cusp and observed as the southern end of the arc,
while the portion within the caustic is observed as the northern
end. We consider the northern part as two merging segments (and call
these c1 and c1), and consider the southern part (which we call c3?)
as a singly imaged arclet.
{\vskip 10pt}
$\bullet$ {\it The `d' system ($z=1.1\pm0.3$):}\\ This is a faint thin
arc, known as \#730, with a number of bright knots visible. Three
elongated components (which we call d1, d2, and d3) can be easily
distinguished along the arc which characterizes it as a cusp arc.

{\vskip 10pt}
\subsection{Secure single distorted arclets:}
Ebbels \etal (1998) have spectroscopically measured the redshifts of a
number of arclets with different degrees of accuracy in the HST/WFPC2
area. We use all the arclets for which they have quality-1 redshifts,
and arclets with quality-2 redshifts in regions where quality-1
redshifts are not available.  Table 2 gives details of the 18 arclets
we use. The last two arclets constitute parts of multiple image
systems.

{\vskip 10pt}
\placetable{tbl-2}
{\small
\begin{tabular}{crrrrr}\hline
ID & $z$\hfil\null & $x$\hfil\null & $y$\hfil\null  & $\theta$\hfil\null 
   & $\epsilon$\hfil\null \\
\tableline \tableline
145       & 0.628&  $ -2.5$&   $103.9$&    26.57&   2.5\\     
159       & 0.564&  $  5.1$&   $105.4$&    80.54&   2.0\\    
158       & 0.723&  $-41.8$&   $ 62.9$&    19.65&   2.2\\   
242       & 0.635&  $ 30.3$&   $ 88.8$&    117.9&   5.0\\    
231       & 0.563&  $ -9.7$&   $ 57.8$&    29.48&   4.0\\    
381       & 0.521&  $ 35.9$&   $ 34.6$&    157.4&   2.5\\   
317       & 0.474&  $ -7.3$&   $ 27.5$&    167.7&   1.5\\  
306       & 0.450&  $-97.5$&   $-44.8$&    135.0&   1.8\\     
464       & 0.476&  $ 10.6$&   $-23.3$&    45.0&    6.0\\        
467       & 0.475&  $ 17.0$&   $-19.7$&    48.01&   4.0\\        
297       & 0.450&  $-21.7$&   $ 23.2$&    65.22&   5.0\\        
456       & 0.538&  $ 20.6$&   $-11.6$&    49.76&   3.5\\     
238       & 0.635&  $-44.4$&   $ 20.1$&    49.40&   4.0\\        
273       & 0.800&  $-41.4$&   $ 16.6$&    68.75&   2.5\\    
205       & 0.693&  $-72.1$&   $ 16.3$&    90.00&   1.2\\     
431       & 0.675&  $ 39.9$&   $ 21.9$&    110.0&   6.0\\        
444       & 1.030&  $ 24.6$&   $ -2.8$&    75.96&   7.0\\            
359 (b3?) & 0.702&  $-20.5$&   $ -3.7$&    98.75&   5.0\\
289 (c3?) & 1.034&  $-63.7$&   $ -8.9$&    77.20&   6.0\\
\tableline 
\end{tabular}
Table 2: Observational parameters of the arclets used in our
reconstruction. The positions $x$ and $y$ are in
arcsec and the position angle $\theta$ is in degrees, while $\epsilon$
is a lower bound on the elongation.  The last two arclets, b3? and c3?
are probably parts of the corresponding multiple image systems in
Table~1, but we have not used these as multiple-image constraints.
We have taken all the arclets as minima of the arrival time, except
for b3? which is evidently a saddle.
{\vskip 10pt}
}

\subsection{Luminosity Distribution}
The apparent magnitudes in the $R$-band of the foreground galaxies are
taken from Le Borgne \etal (1992). We considered only those with
apparent magnitudes $\leq 20.$ and we find that there are 33 of them
enclosed within the HST image. The pixellated light distribution
$L_{mn}$ is obtained by replacing each of the 33 galaxies by a
Gaussian light profile of dispersion $10\arcsec$. Fig.~3
shows a contour plot of this smoothed luminosity distribution.

\figureps[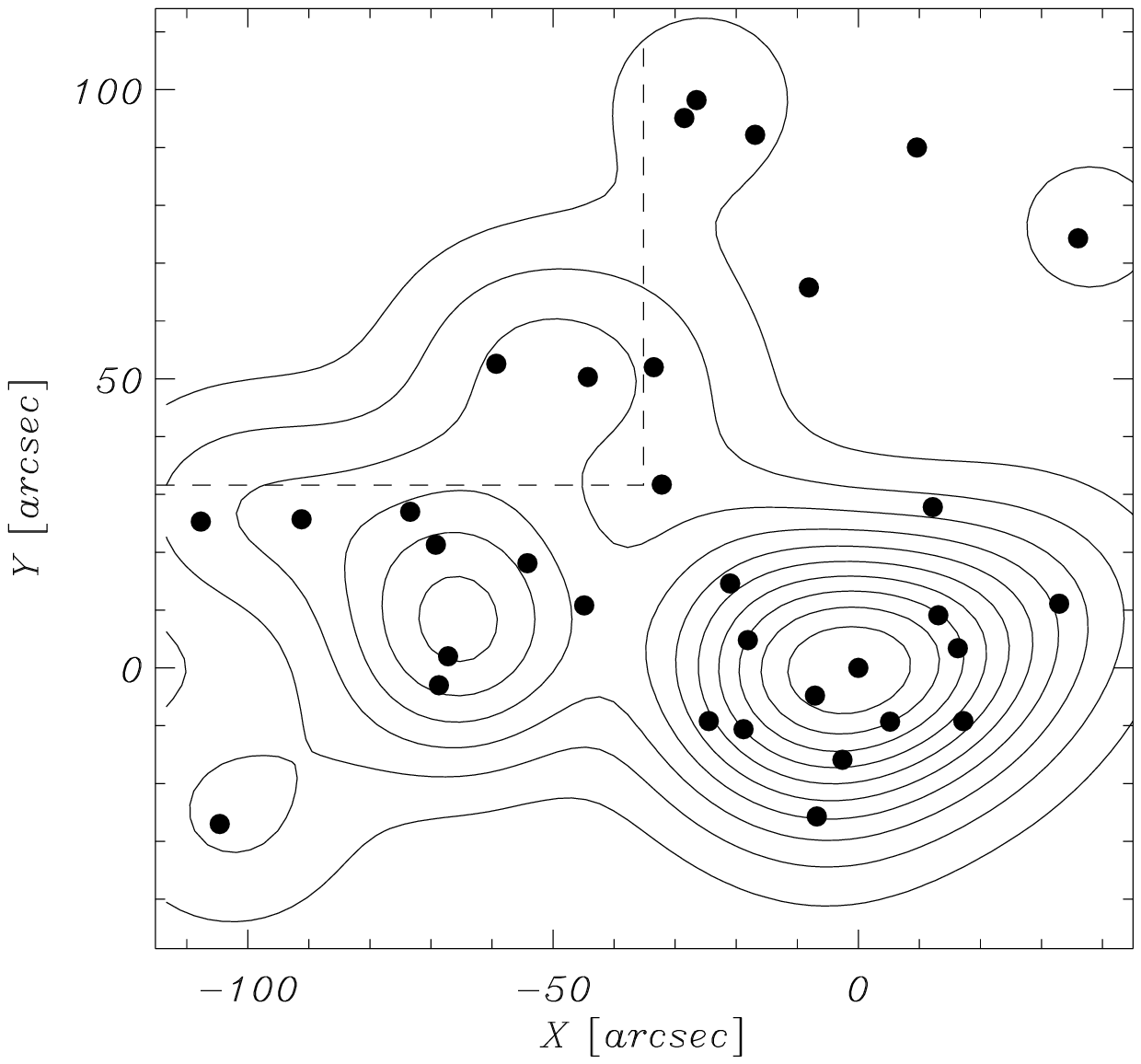,\hsize] 3. The luminosity distribution of the
HST/WFPC2 region of A2218. The positions of the 33 galaxies used are
marked with filled circle.

{\vskip 10pt}
\section{Reconstructed Mass distributions and Uncertainties}

Our optimal mass map of A2218 is shown in Fig.~4.  It uses
a pixel size of $2.29\arcsec$ and regularizes with respect to the
light distribution with smoothing scale $\sigma=6\arcsec$, while
strictly obeying all the strong and weak lensing constraints.  (We
found $\sigma=6\arcsec$ to be the smallest value that eliminated
obvious artifacts in the mass map.  We use this value throughout this
paper except where noted otherwise.)  Our mass map is broadly similar
to the previous parametric models (K95, KESCS96), which show that the
cluster is clearly bimodal with one clump being less massive, and with
detailed substructures implied by the lensing data.

\figureps[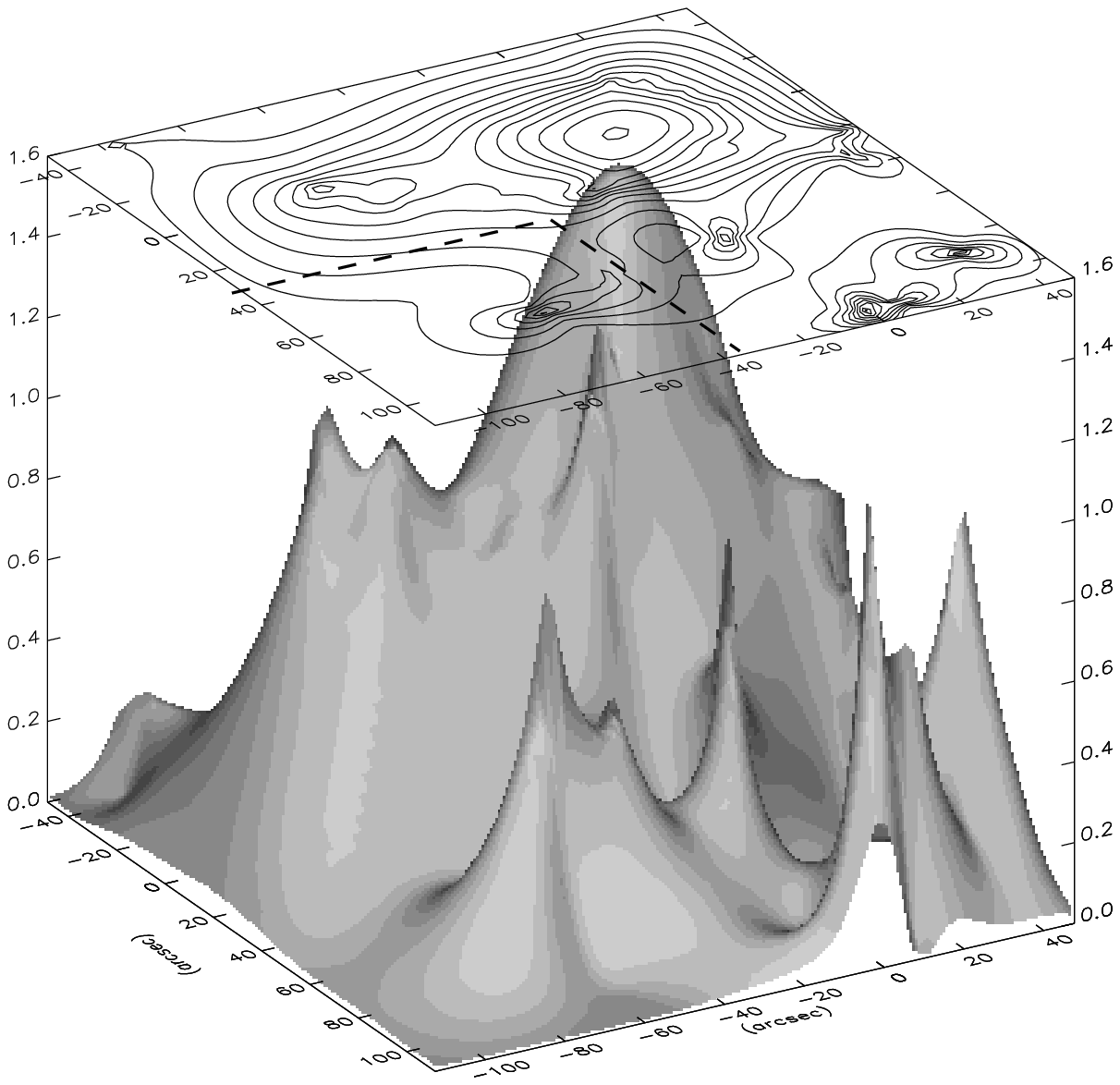,\hsize] 4. Mass distribution of Abell 2218,
reconstructed using strong and weak lensing constraints, and
regularizing with respect to the light distribution and a smoothing
scale of $6\arcsec$.  The density is in units of the critical density
for sources at infinity, $\Sigma_{\rm crit}=3.1\times
10^{14}h_{50}^{-1}\msolar\,\rm arcsec^{-2}$; contours are in steps of
0.1.  Later figures in this paper always use the same units, and use
the same contour steps unless noted otherwise.  The total mass in the
field is $2.75\times 10^{14} h_{50}^{-1}M_\odot$.

It is of interest to examine also other reconstructed mass maps
obtained using different criteria. For example, one can reconstruct a
mass map by assuming zero light distribution for the cluster (i.e.,
setting $L_{mn}=0$ in Eq.\ \ref{reg}); this is shown in Fig.~5, and is
roughly the minimal matter distribution needed to reproduce the data.
Or one can reconstruct a mass map with a much higher smoothing
parameter; Fig.~6 shows a mass map using
$\sigma=19\arcsec$. It is reassuring to see that all the main features
are still present in these cases.

\figureps[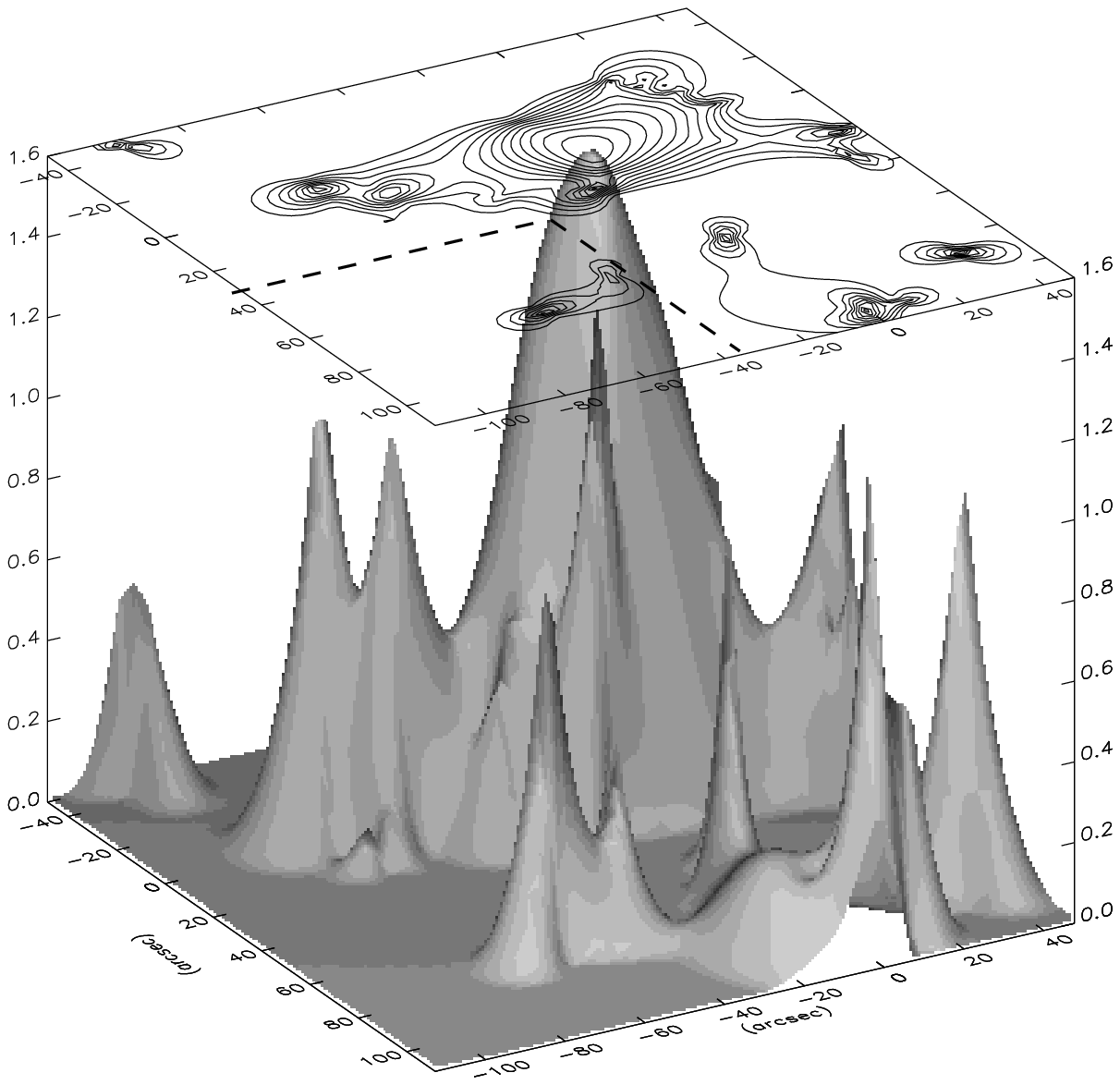,\hsize] 5. Mass distribution reconstructed for
the zero-light case. The total mass in this case is $1.06\times
10^{14}h_{50}^{-1}M_\odot$.

\figureps[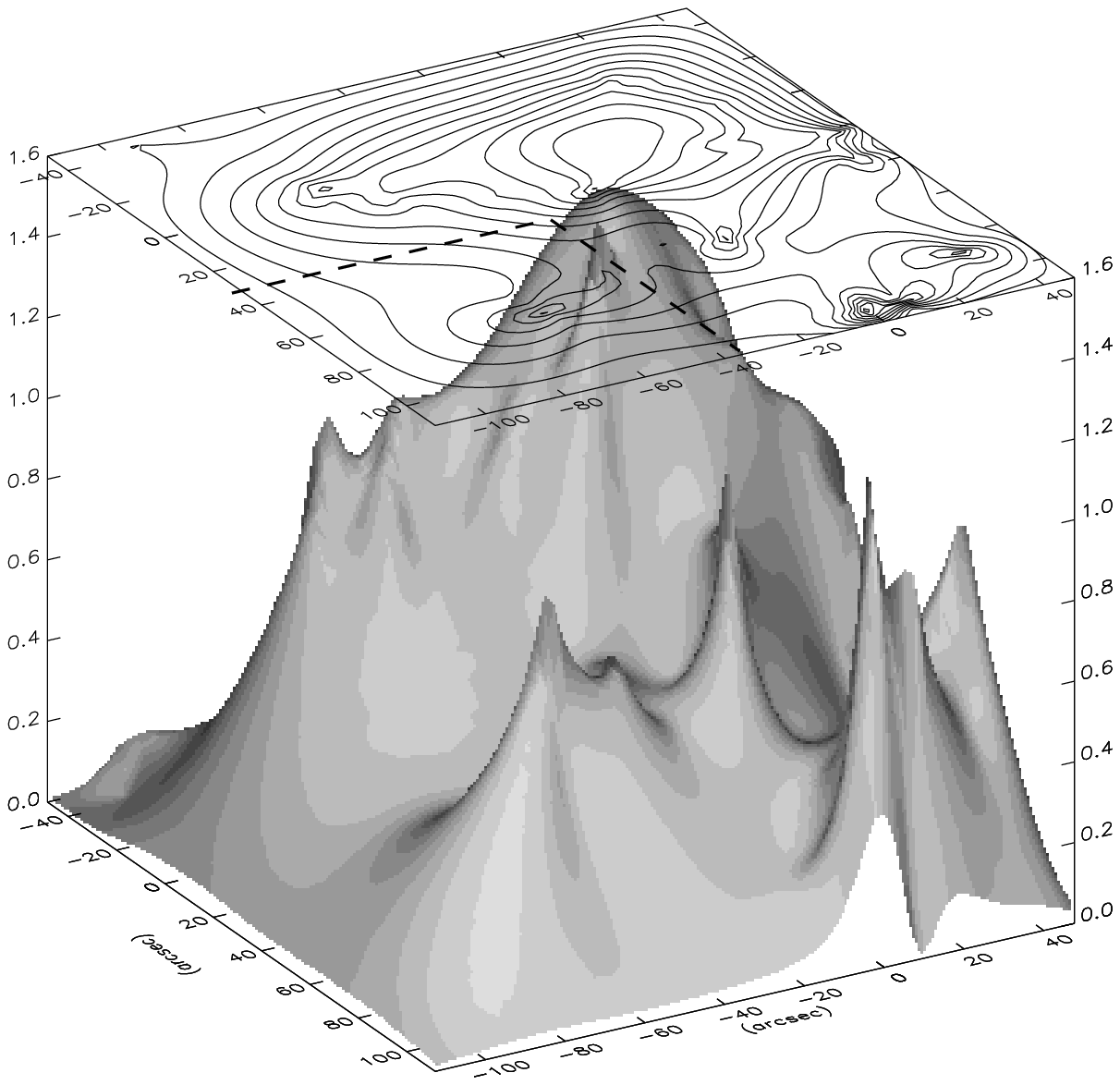,\hsize] 6. Mass map using much more smoothing:
$\sigma=19\arcsec$.

The spikes visible in all our mass maps are due to fact we are using
local constraints from sparsely sampled data, i.e., arclets.  The
peaks of the spikes are very robust between different reconstructions,
but the wings are very variable.  As a result the total mass shows
large variation, though the allowed variation is quite well bounded
above and below.  Statistical distortion maps going to larger radii
would, we expect, constrain the total mass much better and hence
reduce this problem.  We leave this extension for future work,
hopefully with several HST pointings.

\figureps[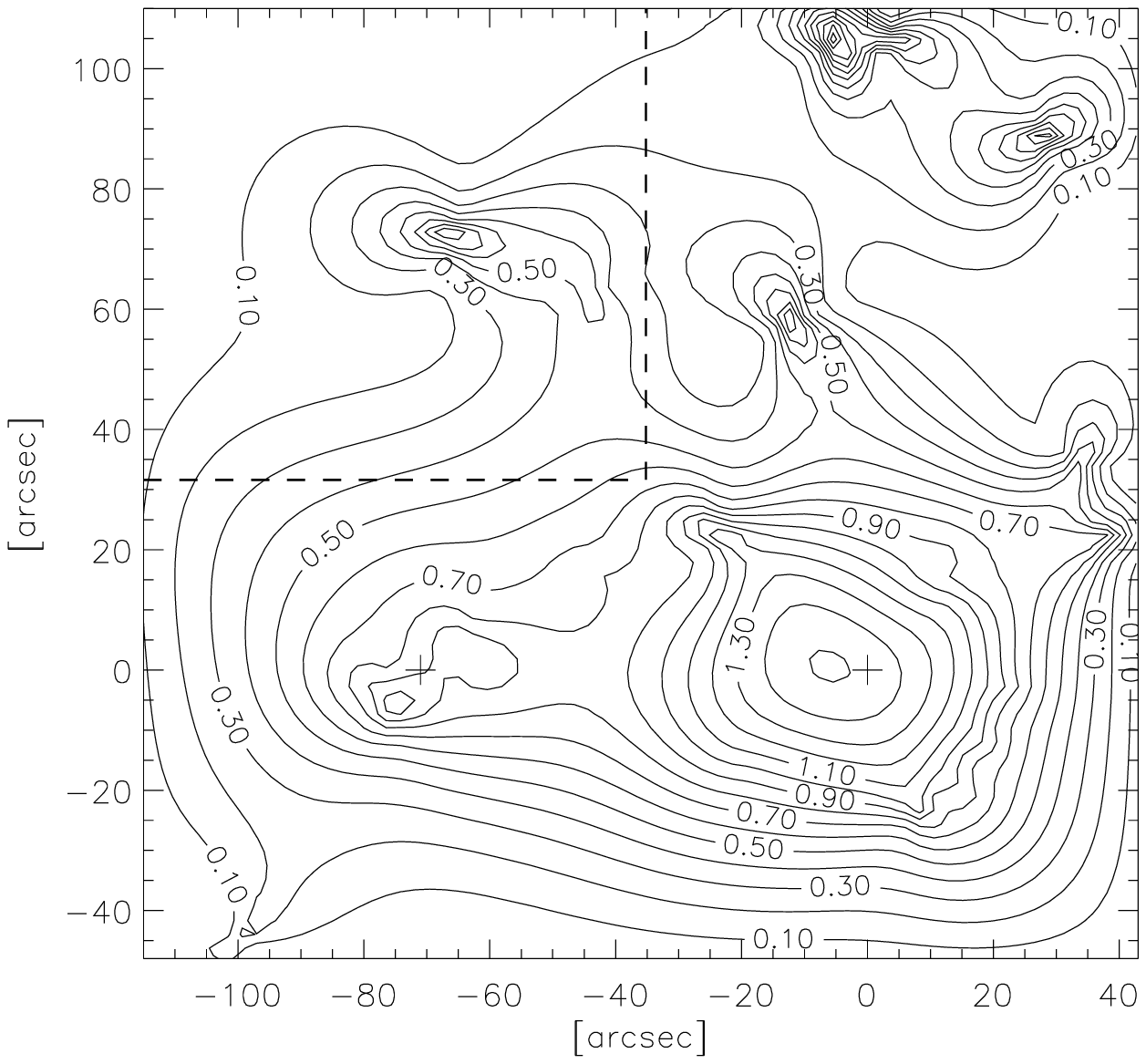,\hsize] 7. Contour map of the projected mass
distribution.  This corresponds to Fig.~4 and is plotted
here for comparison with Fig.~8.

\figureps[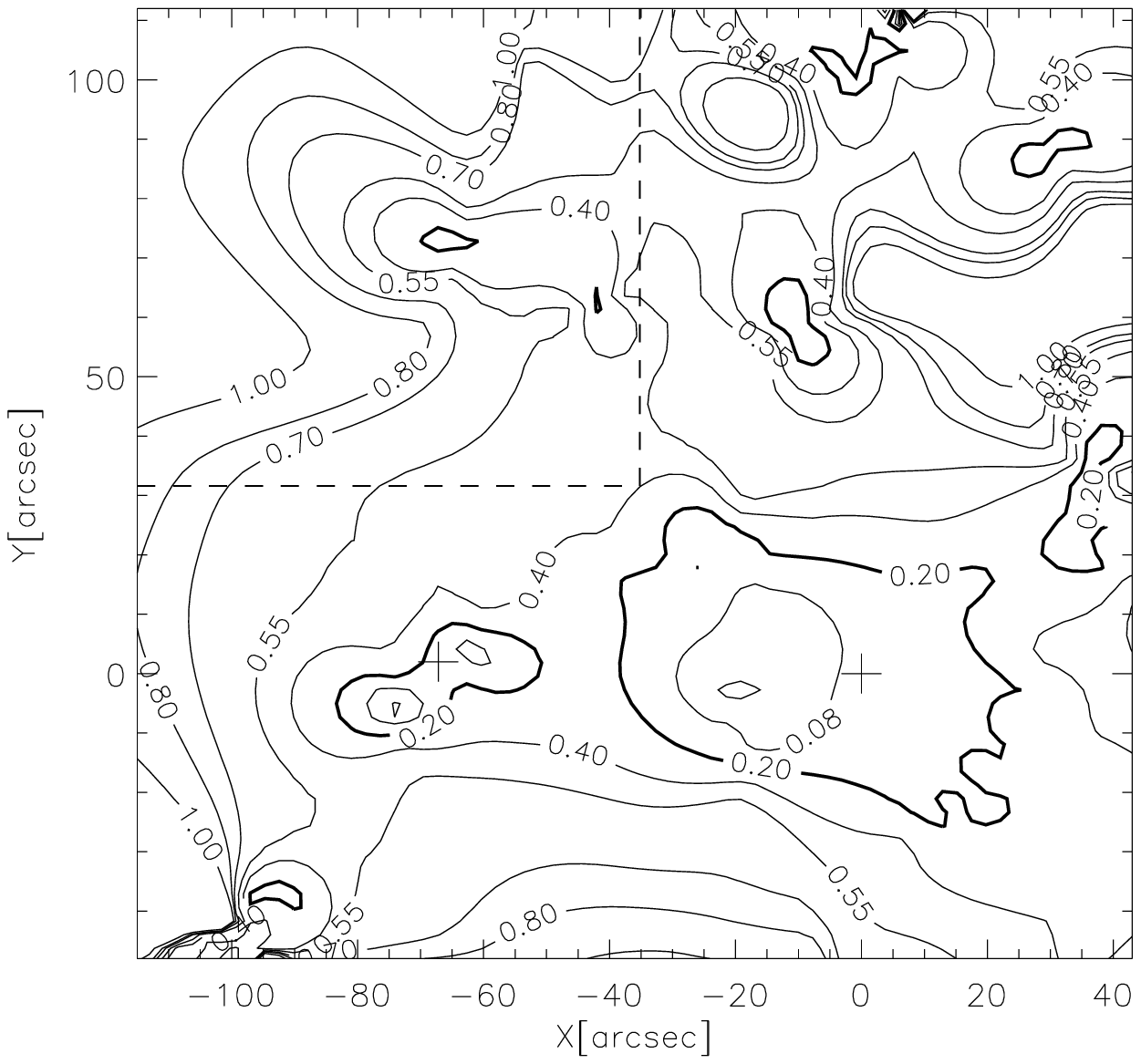,\hsize] 8. Fractional uncertainty of the mass
distribution. The crosses mark the position of the two dominant cD
galaxies in A2218.

Our reconstruction technique naturally lends itself towards
calculating error estimates on the pixellated mass distribution. We
use a Monte-Carlo procedure, randomizing $L_{mn}$ by re-shuffling the
positions of the cluster galaxies or by rotating the entire light
distribution by various angles, and constructing an ensemble
of new mass maps.  The fractional dispersion
\begin{equation}
\Delta\kappa_{mn}=\left[\frac{\langle\kappa_{mn}^2\rangle}
{\langle\kappa_{mn}\rangle^2}-1\right]^{\half}.
\label{rms}
\end{equation}
over this ensemble is a measure of the pixel-by-pixel fractional
uncertainty in the mass map.  This is shown in Fig.~8 as a
contour plot of the computed $\Delta\kappa_{mn}$. Regions in the lens
plane enclosed by $\Delta\kappa_{mn}\le 0.2$, are very well
constrained.  Clearly, the best constrained regions have a high number
density of images.  In Fig.~9 we quantify this by plotting
$\Delta\kappa_{mn}$ binned over circles of radius $10\arcsec$ versus
the the number of images enclosed. As expected, the fractional
uncertainty decreases with the number of images.

\figureps[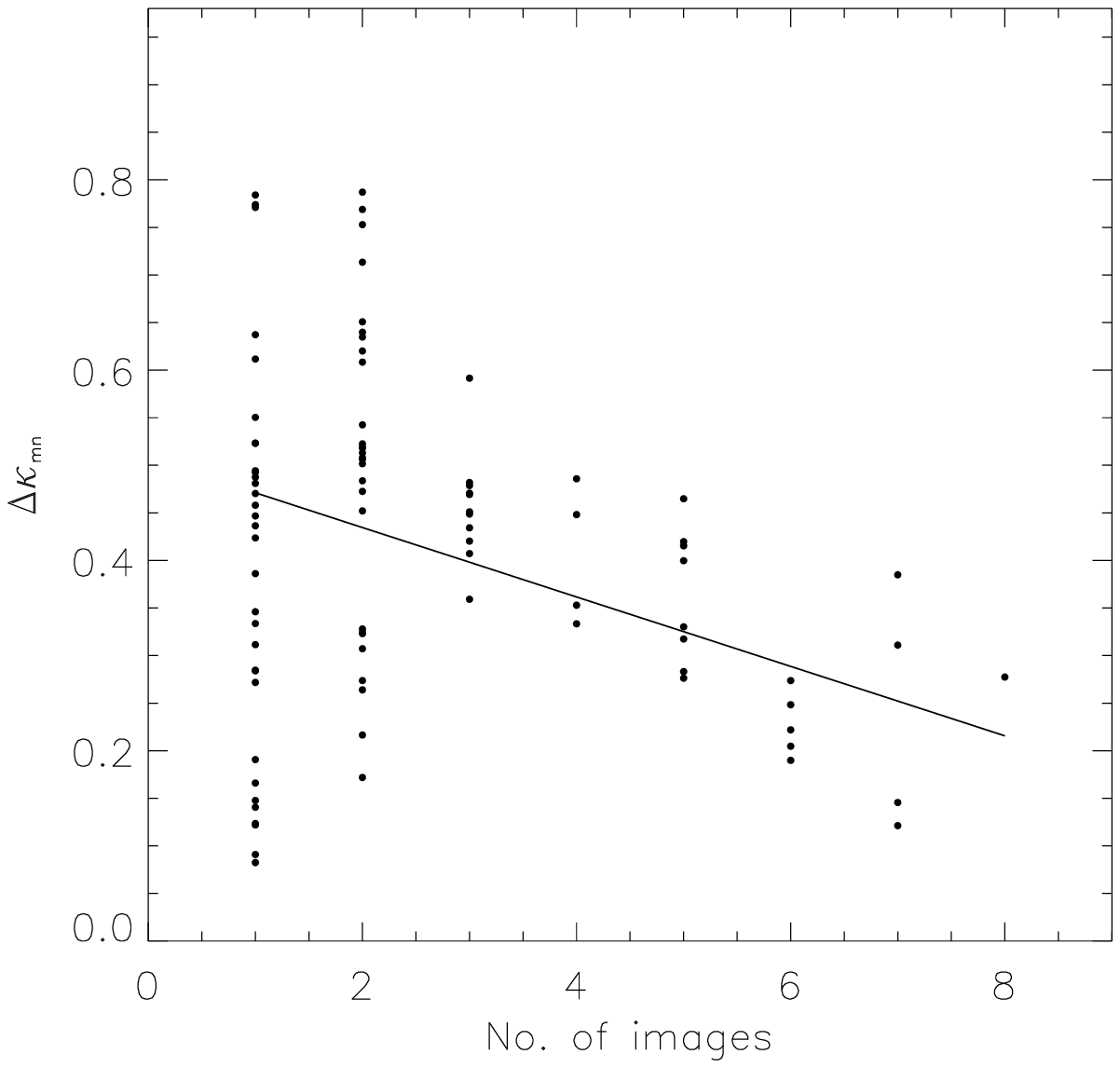,\hsize] 9. The fractional uncertainty versus
the number of images. Each dot represents a circle of radius
$10\arcsec$ over which $\Delta\kappa_{mn}$ and the number of images
have been averaged.  The dashed curve is a linear fit.

\section{Mass-light Discrepancies}

While our mass reconstruction broadly resembles previous parametric
models, we find significant differences that contradict a basic
assumption of the parametric modelling method, that of modelling the
cluster as a smooth mass distribution plus small clumps associated
with bright galaxies.  Two of these differences are worth discussing
in detail.

First, we find that the highest mass peak is significantly offset from
the brightest light peak, the center of the cD galaxy \#391.  The
offset is highly significant given the uncertainties.  To test this
further, we did a mass reconstruction in a smaller field around this
main peak but with a finer pixel size of $1\arcsec$.  We find (see
Fig.~10) that the offset is still persistent and is about
$10\arcsec$ towards the second peak of the cluster. It is interesting
to note that the direction of the offset between the mass and light
peaks detected from our modelling is similar to that found by
Markevitch (1997) between the temperature and light peak.

\figureps[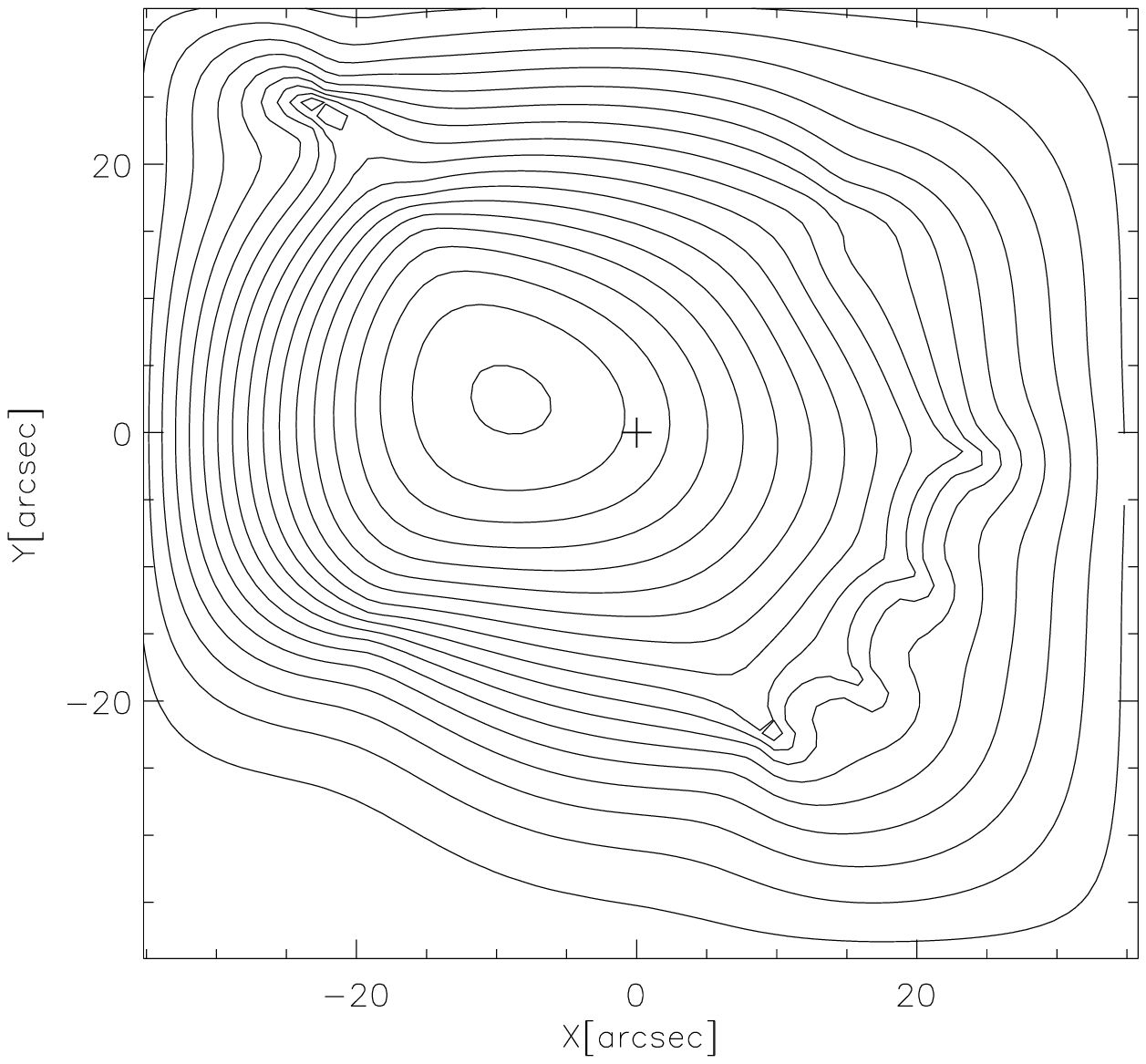,\hsize] 10. Mass reconstruction of a smaller
field, centered on the cD galaxy \#391, with $1\arcsec$ pixels; a
cross marks the cD center.

\figureps[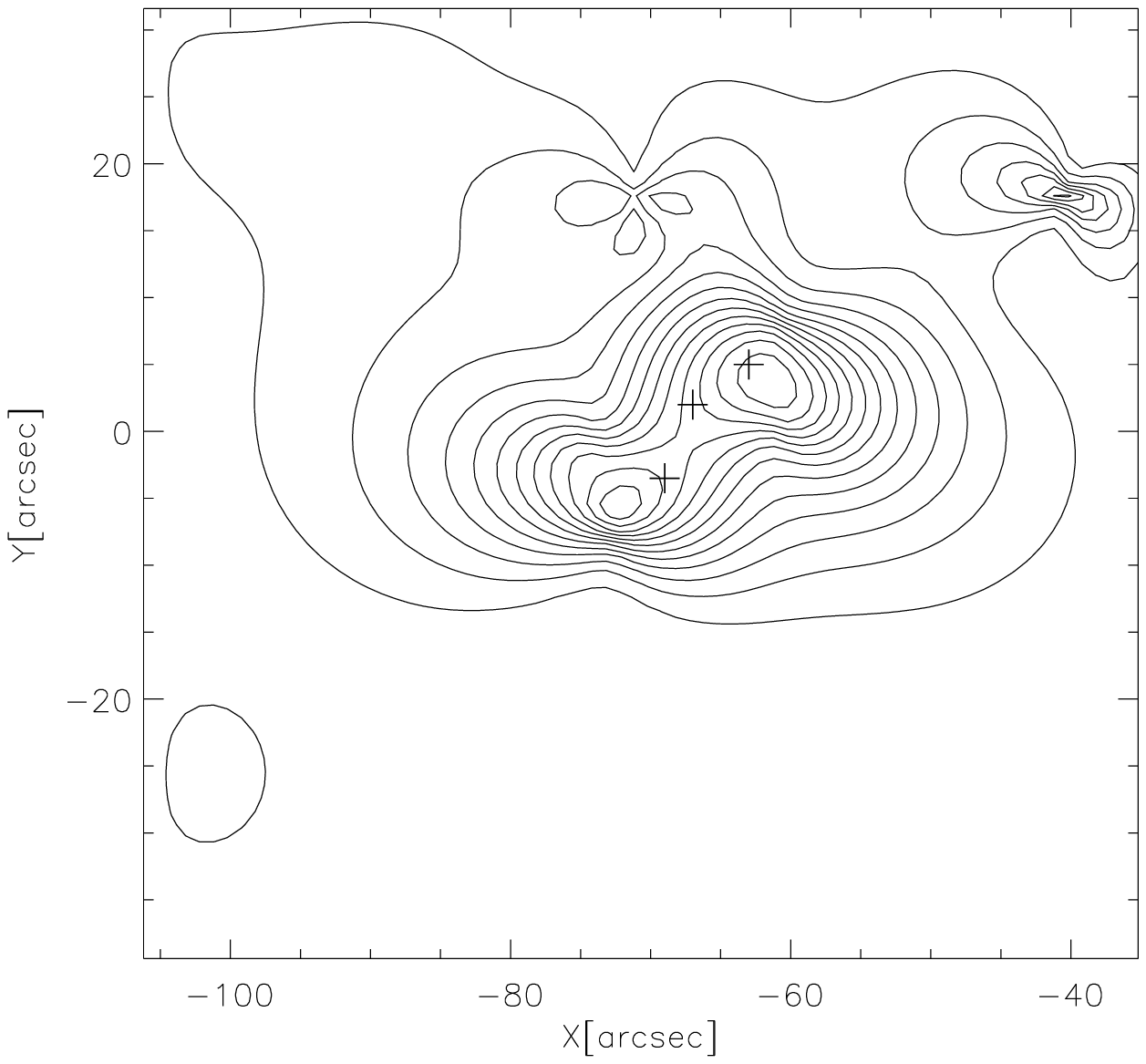,\hsize] 11. Mass reconstruction of a smaller
field centered on the galaxy \#244, with $1\arcsec$ pixels; crosses
mark this galaxy, and two other fainter galaxies.
{\vskip 10pt}

Second, we find that the secondary mass peak is not centered on the
bright galaxy \#244.  Again, we did a mass reconstruction on a smaller
field using $1\arcsec$ pixels, this time near \#244; this is shown in
Fig.\ 11.  The mass distribution in this region, has two
peaks in the directions of two galaxies at least 2.5 mag fainter in
$R$ than \#244, while \#244 itself is in a valley between these two
peaks.  This is also highly significant given the uncertainties, and
is seen both in the main reconstruction and in the `zoom-in'.  If the
three mass concentrations are indeed associated with the identified
galaxies, the $2.5$ mag difference implies that the associated $M/L$
ratios vary by more than a factor of 10.

These new results we reported are robust to pixel size, shape and to
the light distribution of the cluster.

\section{Comparison with X-ray models}

The X-ray emission by clusters is usually modelled assuming that the
underlying potential of the cluster is spherically symmetric and that
the hot gas is in hydrostatic equilibrium.  It is important to test
these assumptions by comparing with lensing, which does not depend on
these assumptions.  Miralda-Escud\'e \& Babul (1995) argued
that the strong lensing data could not be reconciled with an
equilibrium spherical model for the the X-ray emitting has at the
observed temperature---to agree with lensing the gas temperature would
have to be much higher, or equivalently, lensing required at least
twice as much mass as the gas model.  They suggested that the gas
might be partly supported by turbulent motion or by magnetic fields,
or might be multi-phased.  Another possibility is that the cluster is
still undergoing a merging phase, and hence the gas would not be
expected to trace the gravitational potential.

\figureps[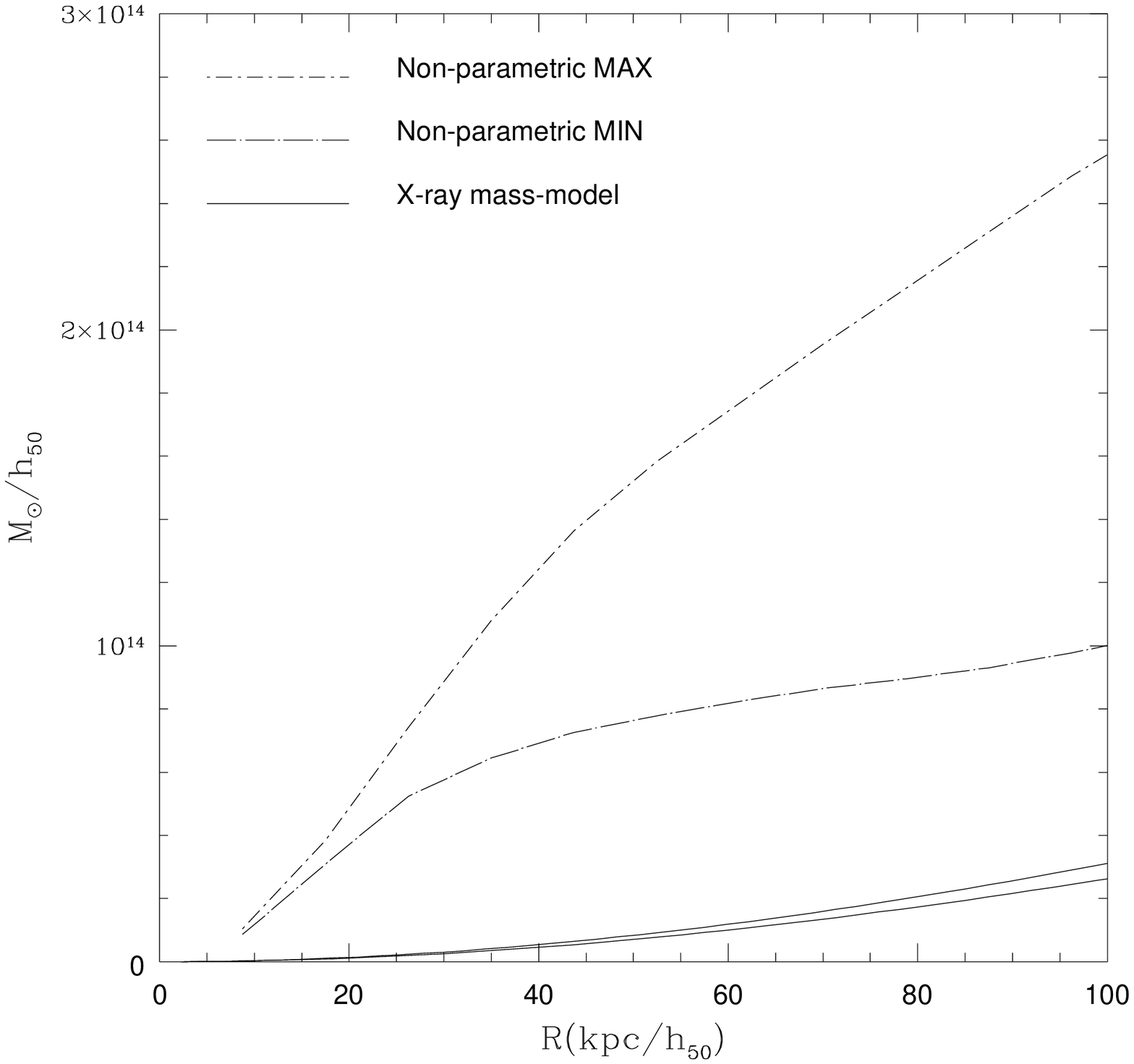,.85\hsize] 12. Projected mass within different
radii from the main cD galaxy \#391, from an X-ray model by Allen et
al. 1998 and from lensing.  The two lensing reconstructions indicated
correspond to Figs 4 (maximum) and 5 (minimum).

In Fig.\ 12 we plot the enclosed projected mass from a model
for the hot gas by Allen (1998) and enclosed mass from our
reconstructions.  The lowest enclosed mass from lensing we got for the
`zero-light' reconstruction shown in Fig.\ 5; the highest
is for our optimal reconstruction (Fig.\ 4) which tends to
minimize $M/L$ variation and thus extrapolates mass into regions with
now lensing data; other models correspond to various other
regularizations.  We see from Fig.\ 12 that the mass
discrepancy is a factor of 2.5 for even the unrealistic-looking model
of Fig.\ 5.

\section{Analysis of individual multiple image systems}

We have examined the arrival time surfaces, critical curves, and
caustics for all the multiple image systems and some of the arclets.
Arrival time contours are a good way of verifying a mass distribution
reproduces the given image position properties.  In this section we
discuss two of the multiple image systems in detail, and report
predicted counter-images, if any.

\subsection{The a1-a2 system}

\figureps[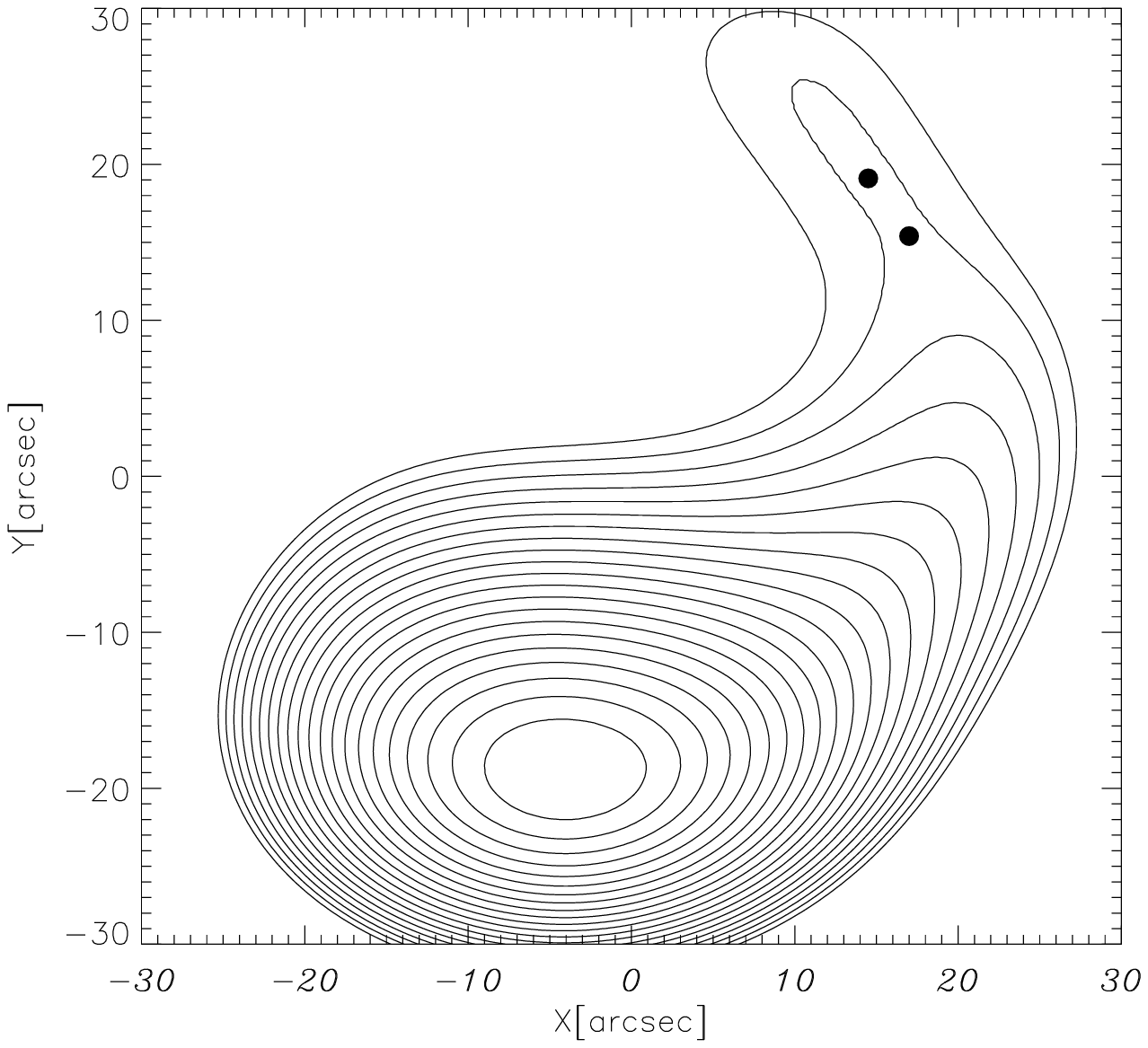,\hsize] 13. Arrival time surface for the a1-a2
system at z=2.515. Filled circles mark the observed positions of a1
and a2.  A third image is predicted roughly at
($-5\arcsec,-23\arcsec$).

The arrival time surface (see Fig.\ 13) reproduces the
image positions of the two segments a1 and a2 comprising the fold arc.
Moreover, the arrival time surface predicts a third image 20$\arcsec$
below the dominant mass clump, very close to the observed arclet
\#468, also predicted by K96 as a counter image for the same arc.
The mirror symmetry seen across the arc a1-a2 implies that the
critical curve is passing through them. Exploring the critical curves
at the redshift ($z=2.515$) of the system confirms that a1-a2 is
indeed a fold arc resulting from a background source lying close to a
beak-to-beak caustic. The critical curves and caustics implied by our
mass map are shown in Fig.\ 14.

\figureps[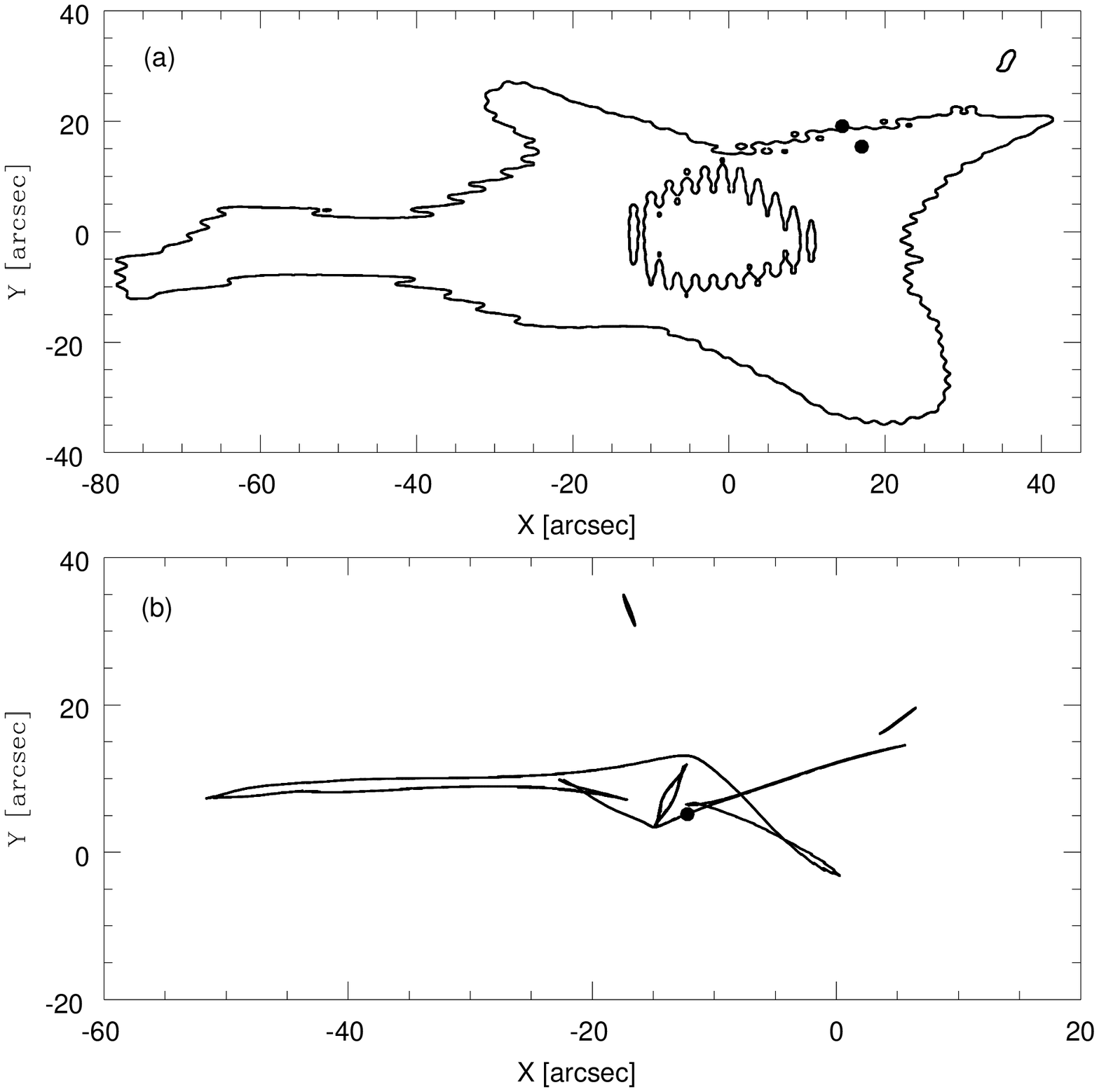,\hsize] 14. The critical curves and caustics at
the redshift of the a1-a2 system ($z=2.515$). The upper panel shows
the critical curves with the two filled circles marking the positions
of the two merging images. The lower panel is the corresponding
caustics with the filled circle marking the predicted source
position.

\subsection{The b1-b2 system}

The arrival time surface (see Fig.\ 15) in addition to
reproducing the positions of b1 and b2 as required, also predicts an
image near the observed position of b3?, but not exactly coinciding
with it, which suggests that b3? is an image of a different component
of the same lens galaxy.  Fig.\ 16 shows the critical curves
caustics at the redshift ($z=0.702$) of this system.  The predicted
position of the source is in a region where the caustics are very
convoluted and this would result in a complex image configuration.  As
an auxiliary test, we inspect the arrival time surface for the arclet
b3? itself (see Fig.~17). The position of the arclet b3?
is of course exactly reproduced and, moreover, four extra images are
predicted. The position of two of the predicted images matches fairly
with those of the b1, and b2. This result strongly supports the
scenario that the images b1, b2 and \#337 are counter images of the
arclet b3?.

\figureps[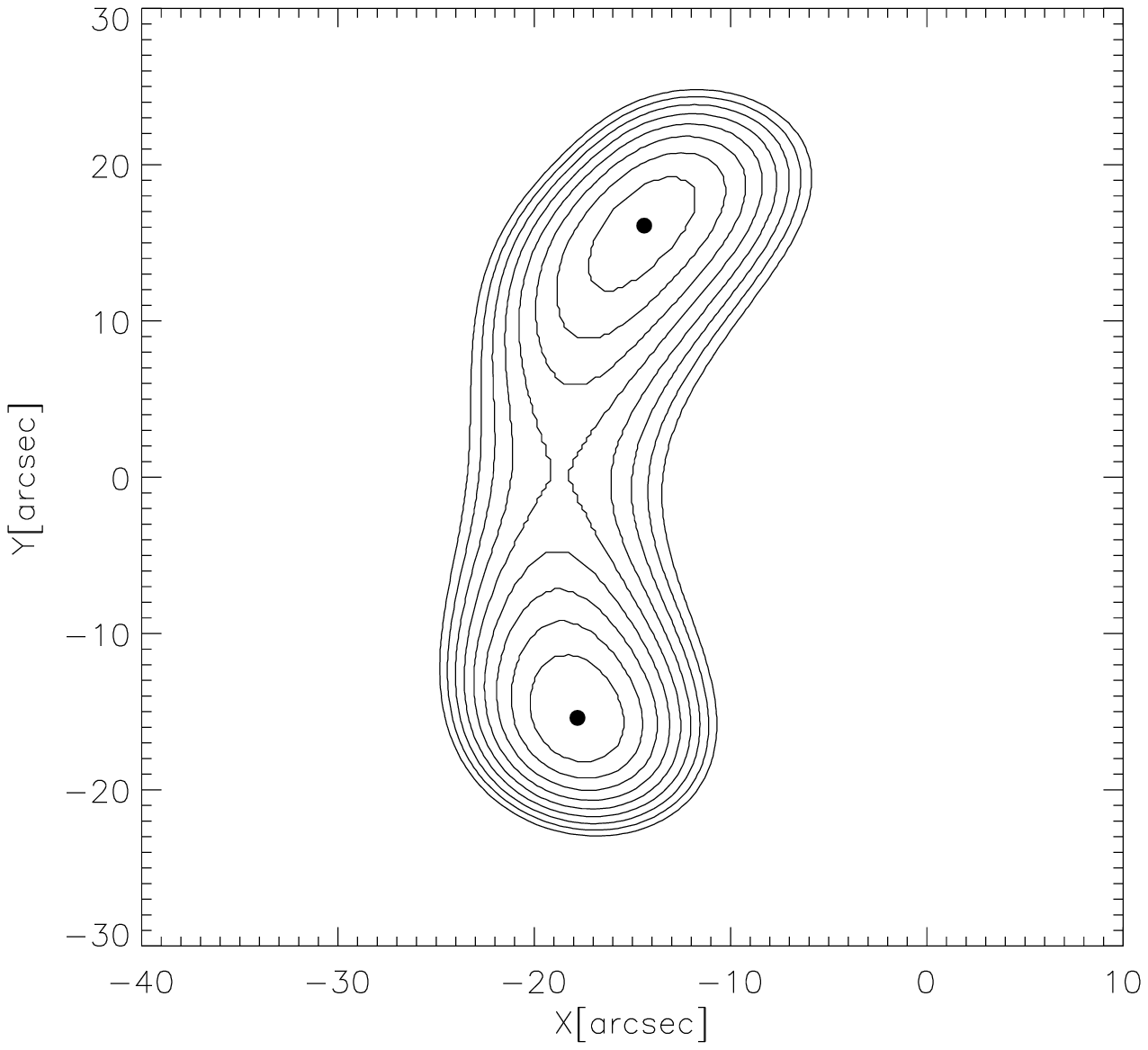,\hsize] 15. Arrival time surface for the b1-b2
system at z=0.702. Filled circles mark the observed positions of b1
and b2.

\figureps[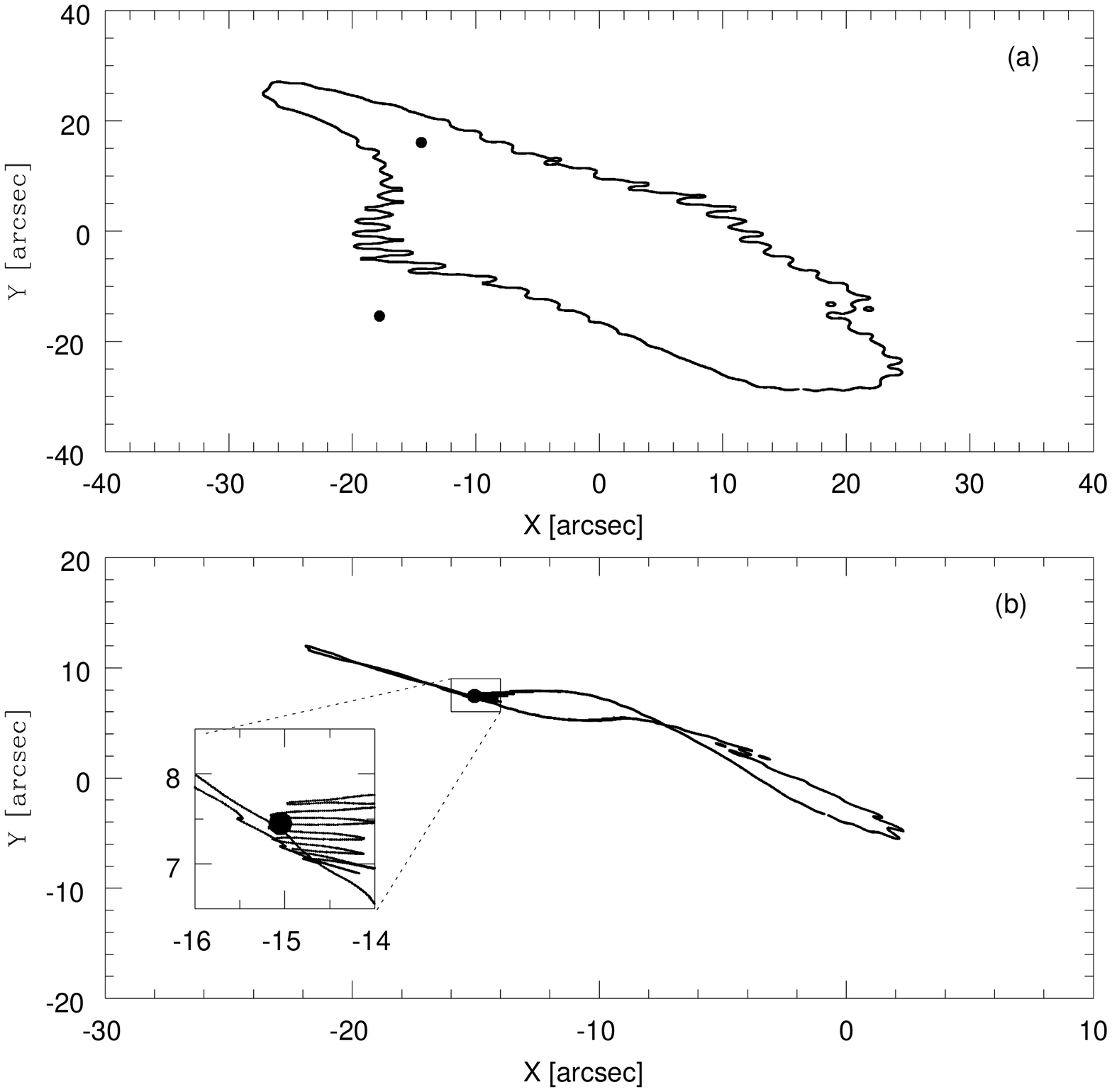,\hsize] 16. The critical curves and caustics at
redshift $z=0.702$. The critical curve (upper panel) with filled
circles marking positions of observed images. Caustics (lower panel)
with filled circle marking the predicted position of the source.

\figureps[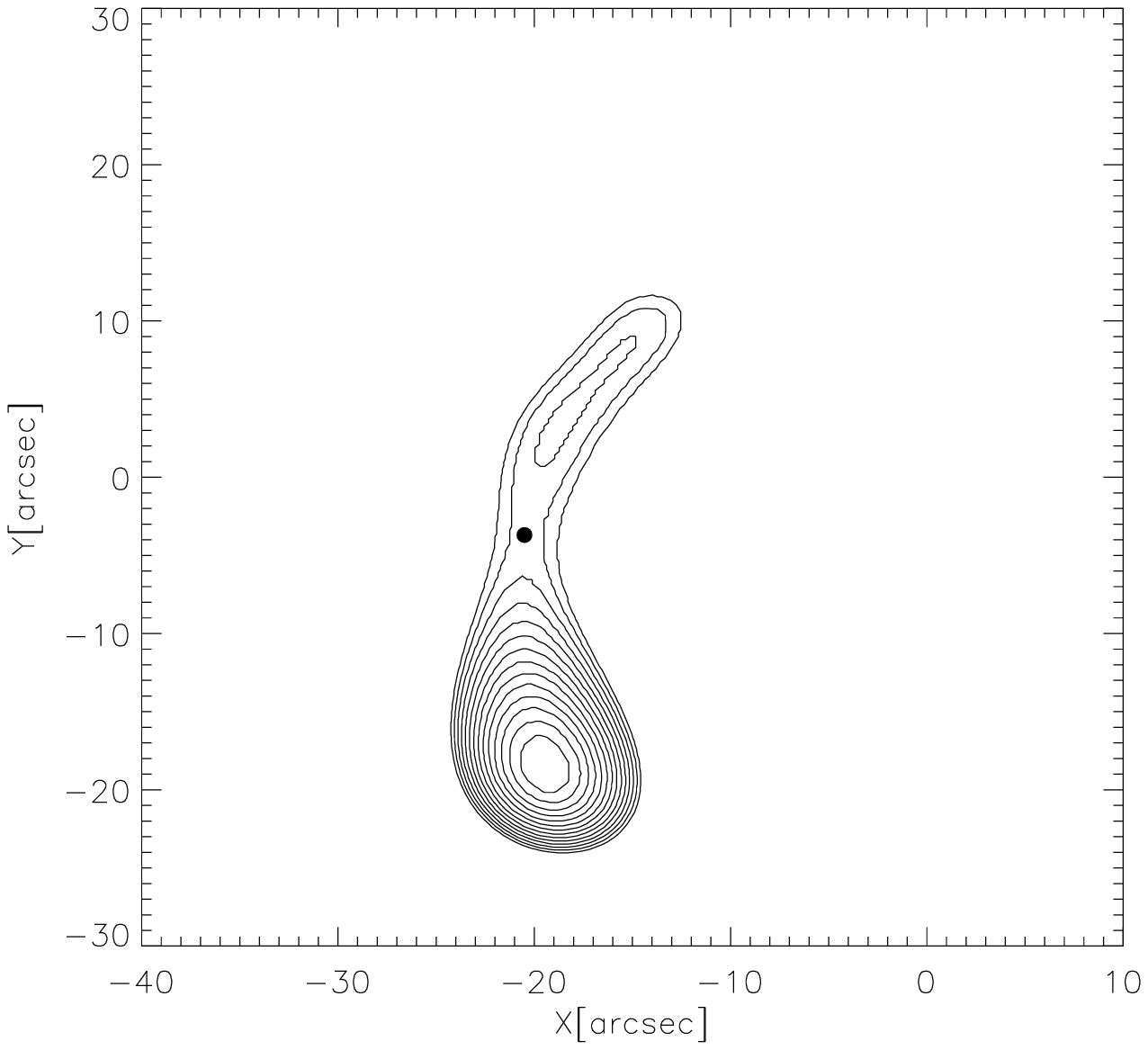,\hsize] 17. Arrival time surface for b3? at
z=0.702. The filled circle marks the observed
position of b3?.

\subsection{Statistical Magnification Map}

Since most work on cluster lensing at present is based on statistical
distortion of images, it is interesting to examine the statistical
magnification map.  We calculate the statistical magnification as
follows.  We first divide the field into square regions of side
$10\arcsec$ or $20\arcsec$.  Within each square we compute the
magnification matrix at 25 random points, using our optimal mass
reconstruction and redshifts randomly chosen from those of the
multiply imaged systems; of these we discard any points corresponding
to axis ratios $>5$ and average the rest.  This averaged matrix we
call the statistical magnification.  Such a procedure roughly mimics
the observational procedure of averaging ellipticities while
discarding obvious arclets, though of course it is not the same
because for real data the absolute magnification is not usually
available.


Fig.\ 18 shows the statistical magnification we obtained
by averaging over $10\arcsec$ and $20\arcsec$ squares.  It illustrates
a cautionary fact: in the strong lensing region the statistical
magnification is dominated by noise.  The reason is that a critical
curve might be crossed and this can change the shear from $\sim1$ to
$\infty$ and back to $\sim1$ again over a $10\arcsec$ scale.
Discarding arclets with high axis ratios does not cure this
problem---the left panel in Fig.\ 18 shows two places with
statistical magnification corresponding to axis ratios $>5$ even
though individual location with such high axis ratios have been
excluded; this is because two low-axis-ratio magnification matrices on
opposite sides of a critical curve can result in a high-axis-ratio one
if averaged. 

\hfigureps[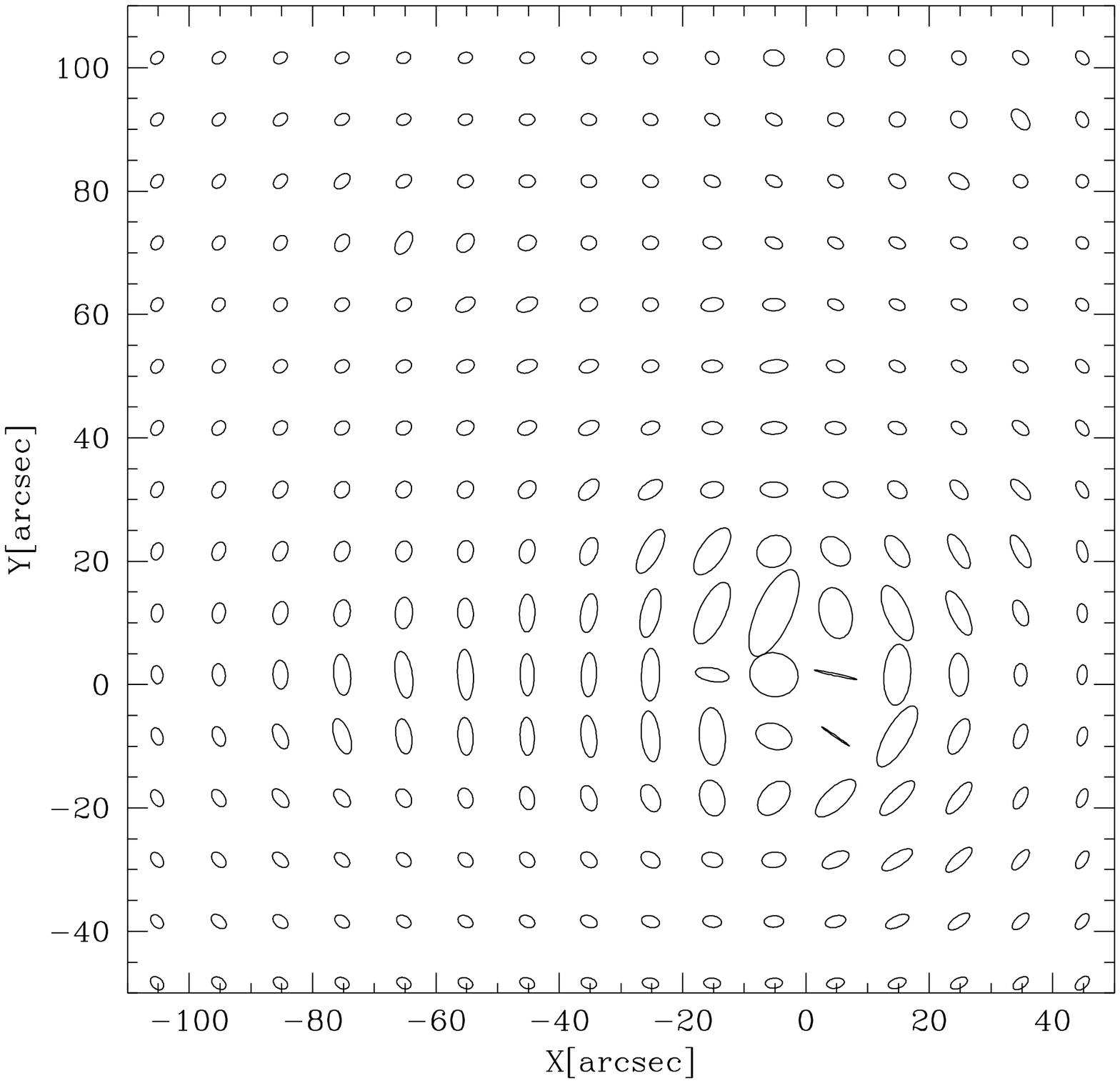,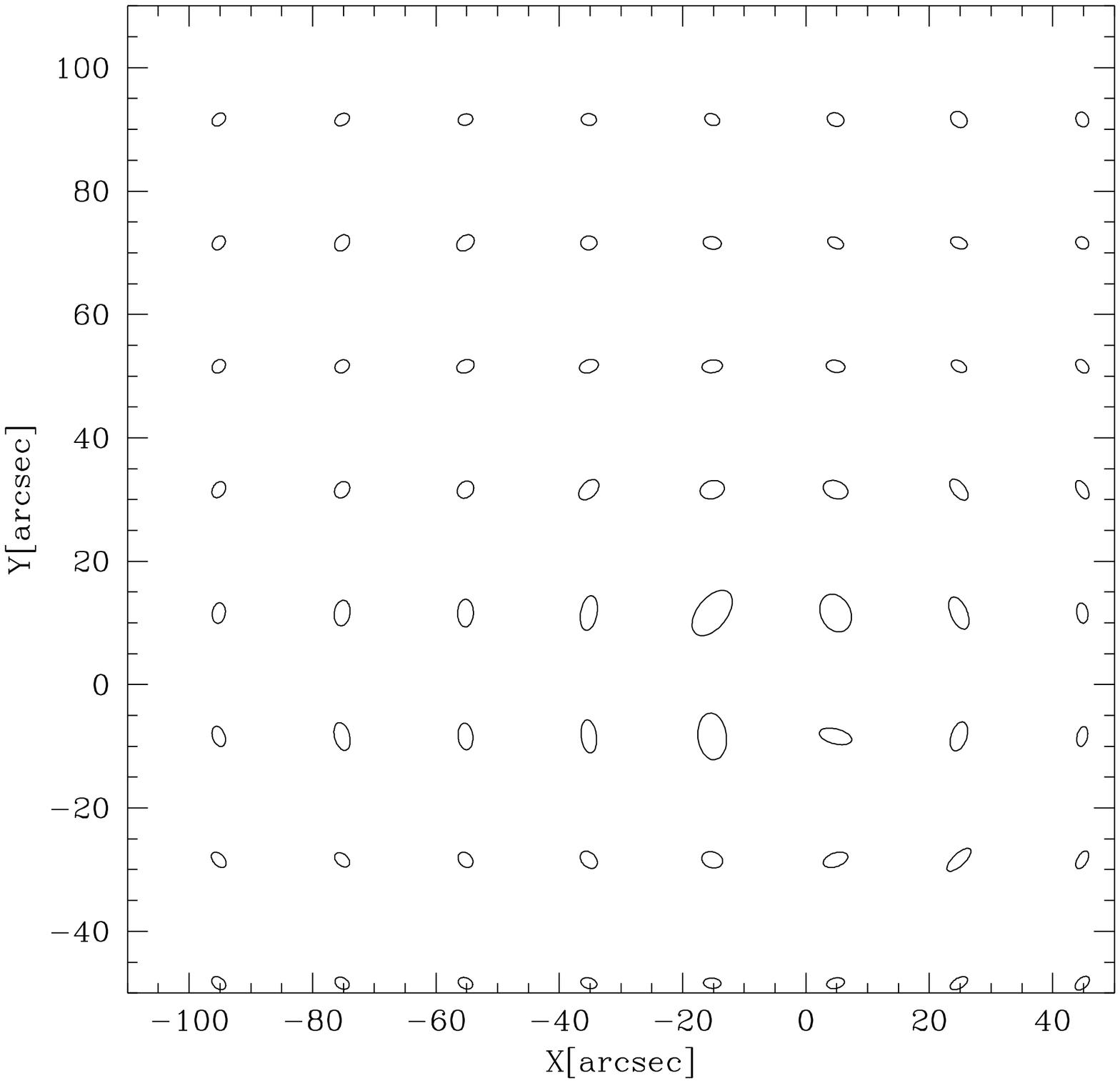,.5\hsize] 18. Statistical
magnification maps from our optimal mass reconstruction, averaging
over $10\arcsec$ squares (top), and $20\arcsec$ squares (bottom).  The
size of each ellipse represents the convergence, the shape represents
the shear.

We conclude that statistical shear must not be used in the strong
lensing region of clusters; instead, constraints from each image
should be used separately.

\section{Conclusions}

The present paper is, to our knowledge, the first to combine
constraints from strong and weak lensing in cluster mass
reconstruction. The method we describe recovers a pixellated mass
distribution that strictly obeys constraints from lensing observations
and uses the light distribution of the cluster as subordinate
information that may be overridden by lensing constraints. We explore
the projected (dark) matter distribution of Abell 2218 with an
unprecedented level of detail. Our mass map shows that the primary
mass peak is offset from the the light peak and that projected
mass-to-light variations of galaxy-sized components are severely
inconsistent with the galaxy $M/L$ scaling deduced from dynamical
arguments. In general, these imply that mass does not follow light on
a range of length scales in A2218. We also confirm and elaborate a
previous result that current X-ray mass models significantly
underestimate the mass of a cluster that is able to reproduce the
observed lensing. Our results conclude that mass estimates from
lensing are {\it at least\/} 2.5 times that from X-ray models. This
suggests that in at least some clusters intra-cluster hot gas does not
trace the gravitational potential.

\section*{Acknowledgements}

The authors would like to thank Steve Allan for providing X-ray mass
model of A2218, and Susan Ridgway for reducing the HST archival
data. We are also grateful to Ramesh Narayan for an insightful
discussion on the mass sheet degeneracy.

HMA acknowledges financial support from the Overseas Research Scheme
(ORS), Oxford Overseas Bursary (OOB) and Wolfson College Bursary. LLRW
would like to acknowledge PPARC postdoctoral fellowship at IoA,
Cambridge.

\appendix 
\section{Integrals over Gaussian Pixels} 

Integrating (\ref{pixint}) yields the coefficient of the $mn$-th
pixel's contribution to the potential:
\begin{equation}
\psi_{mn}(\theta_x,\theta_y) =
\frac{a^2}{4}\left[\ln\left(u\right)- 
{\rm Ei}(-u)+\gamma_E\right],
\qquad u \equiv \frac{2\theta^2}{a^2},
\label{psi}
\end{equation}
where $\rm Ei$ denotes the exponential integral and $\gamma_E$ is
Euler's constant. Thus the first derivatives of the coefficients of
the deflection potential, i.e., the components of the deflection angle,
are
\begin{eqnarray}
\partial_x\psi_{mn}(\theta_x,\theta_y)
&=& \frac{\theta_x}{u}\left[1-\exp(-u)\right]\nm\\
\partial_y\psi_{mn}(\theta_x,\theta_y)
&=& \frac{\theta_y}{u}\left[1-\exp(-u)\right].\nm\\
\end{eqnarray} 
The second derivatives of the potential are
\begin{eqnarray}
\partial_{xx}\psi_{mn}(\theta_x,\theta_y) &=&
\frac{1}{u}\left[1-\frac{2\theta_x^2}{\theta^2}\right]
\left[1-\exp(-u)\right]+\frac{2\theta_x^2}{\theta^2}\exp(-u),\nm\\
\partial_{yy}\psi_{mn}(\theta_x,\theta_y) &=&
\frac{1}{u}\left[1-\frac{2\theta_y^2}{\theta^2}\right]
\left[1-\exp(-u)\right]+
\frac{2\theta_y^2}{\theta^2}\exp(-u),\nm\\
\partial_{xy}\psi_{mn}(\theta_x,\theta_y) &=&
\frac{2\theta_y\theta_x}{\theta^2}
\left[\exp(-u)-\frac{1}{u}\left[1-\exp(-u)\right]\right].
\label{psixy}
\end{eqnarray}

\end{document}